\documentclass[12pt]{article}

\usepackage{amssymb}
\usepackage{epsfig}
\usepackage{graphics}

\newcommand {\beq} {\begin{eqnarray}}
\newcommand {\eeq} {\end{eqnarray}}

\newcommand {\cose} {\cos \theta_{\tilde e}}
\newcommand {\sine} {\sin \theta_{\tilde e}}
\newcommand {\cosesq} {\cos^2 \theta_{\tilde e}}
\newcommand {\sinesq} {\sin^2 \theta_{\tilde e}}

\newcommand {\exclusion} [1] {}
\newcommand {\fig} [1] {Fig.~\ref{#1}}

\newcommand {\msel} [2] {m_{ {\tilde e}^{#1}_{#2} } }

\newcommand {\sect} [1] {Sect.~\ref{#1}}
\newcommand {\sel} [2] { {\tilde e}^{#1}_{#2} }
\newcommand {\selm} [1] { {\tilde e}^{-}_{#1} }
\newcommand {\selp} [1] { {\tilde e}^{+}_{#1} }
\newcommand {\tanb}  {\tan \beta}

\def\lsim{\raise0.3ex\hbox{$\;<$\kern-0.75em\raise-1.1ex\hbox{$\sim\;$}}}
\def\gsim{\raise0.3ex\hbox{$\;>$\kern-0.75em\raise-1.1ex\hbox{$\sim\;$}}}

\usepackage{latexsym}
\usepackage[dvips]{color}
\def\greaterthansquiggle{\raise.3ex\hbox{$>$\kern-.75em\lower1ex\hbox{$\sim$}}}
\def\lessthansquiggle{\raise.3ex\hbox{$<$\kern-.75em\lower1ex\hbox{$\sim$}}}

\newcommand{\beqn}{\begin{eqnarray}}
\newcommand{\eeqn}{\end{eqnarray}}
\newcommand{\bequ}{\begin{equation}}
\newcommand{\eequ}{\end{equation}}
\newcommand{\bsl}{\begin{sloppypar}}
\newcommand{\esl}{\end{sloppypar}}

\begin{document}
\title{ {\flushright{\small WUE-ITP-2002.002 \\
                    \small DESY 01-173 \\
                    \small HEPHY-PUB 743 \\
                    \small ZU-TH 38/01 \\ .}}\\
       Selectron Pair Production at $e^-e^-$ and $e^+e^-$ Colliders \\with 
        Polarized Beams}
\author{C. Bl\"ochinger$^1$, H. Fraas$^1$, G. Moortgat-Pick$^2$
        and W. Porod$^{3,4}$ \\[0.3cm]
    \small $^1$Institut f\"ur Theoretische Physik und Astrophysik, \\\small
            Universit\"at W\"urzburg, D-97074 W\"urzburg, Germany \\\small
    $^2$DESY, Deutsches Elektronen-Synchrotron, D-22603 Hamburg, Germany \\\small
    $^3$Inst.~f.~Hochenergiephysik, \"Oster.~Akademie d.~Wissenschaften,\\\small
      A-1050 Vienna, Austria \\\small
    $^4$Inst.~f\"ur Theor. Physik, Universit\"at Z\"urich, CH-8057 Z\"urich,
      Switzerland
}                     
%
%
%
%
\maketitle

\begin{abstract}
 We investigate selectron pair production and decay in
  $e^-e^-$ scattering and $e^+e^-$ annihilation with polarized beams
  taking into account neutralino mixing as well as ISR and
  beamstrahlung corrections.  One of the main advantages of having
  both modes at disposal is their complementarity concerning the
  threshold behaviour of selectron pair production. In $e^-e^-$ the
  cross sections at threshold for $\tilde{e}_R \tilde{e}_R$ and
  $\tilde{e}_L \tilde{e}_L$ rise proportional to the momentum of the
  selectron and in $e^+ e^-$ that for $\tilde{e}_R \tilde{e}_L$.
  Measurements at threshold with polarized beams can be used to
  determine the selectron masses $m_{\tilde{e}_{L/R}}$ precisely.
  Moreover we discuss how polarized electron and positron beams can be
  used to establish directly the weak quantum numbers of the
  selectrons.  We also use selectron pair production to determine the
  gaugino mass parameter $M_1$. This is of particular interest for
  scenarios with non-universal gaugino masses at a high scale
  resulting in $|M_1| \gg |M_2|$ at the electroweak scale.  Moreover,
  we consider also the case of a non-vanishing selectron mixing and
  demonstrate that it leads to a significant change in the
  phenomenology of selectrons.
\end{abstract}

\section{Introduction}
\label{intro}

Supersymmetry (SUSY) is one of the most promising concepts for physics 
beyond the
Standard Model (SM) and we expect that candidates for supersymmetric particles
will first be discovered at the LHC. Due to the clear signatures 
linear colliders are well suited for high precision studies
\cite{EpemOverview,report,TDR}. In addition to
suitable cuts a simultaneous polarization of both beams is crucial
for suppression of the background. One of the
most important goals of a future $e^+e^-$ or $e^-e^-$ linear collider (LC)
 will be the precise determination of 
quantum numbers and couplings of supersymmetric particles
in order to establish the supersymmetric framework.
Also for
some of these measurements beam polarization is indispensable
\cite{herb}.

Since in many cases the $e^+ e^-$ mode is favourable most of the
running time will be spent for this mode. However, for some processes
the $e^- e^-$ mode is accepted to be superior, particularly for the
precise measurement of the selectron masses \cite{Feng98,Peskin1}.  This
is due to the steeper rise of the cross sections for
$\tilde{e}^-_R\tilde{e}^-_R$ and $\tilde{e}^-_L\tilde{e}^-_L$
production in $e^-e^-$ at the threshold compared to 
selectron production in $e^+e^-$ annihilation
\cite{Feng98,Peskin1,Martyn1}.  Most of these analyses assume that the
lightest neutralino is a pure B--ino so that in
$\tilde{e}^-_R\tilde{e}^-_R$ production the contributions from the
exchange of the other neutralinos can be neglected.  In the study
presented here we take into account neutralino mixing and give also
results for $\tilde{e}^-_L \tilde{e}^-_L$ and $\tilde{e}^-_R
\tilde{e}^-_L$ production.  
In the $e^-e^-$ mode the cross sections are in general larger
than in the $e^+ e^-$ mode due to the absence of
destructive interferences between s--channel and t/u--channel
exchange \cite{Bartl87}. Moreover, a significantly 
lower background from SM and SUSY is
expected in the $e^-e^-$ mode. 

In order to establish Supersymmetry 
it is necessary to verify experimentally
the association of chiral fermions and their scalar SUSY partners.
Moreover, for the identification of the supersymmetric scenario the
precise determination of model parameters is necessary. In particular for the
determination of the gaugino mass parameter $M_1$ of the MSSM and
a test of gaugino mass unification divers procedures have already been
discussed \cite{Feng98,Moortgat-Pick:2000uz}.
Since 
the cross section for $e^- e^-\to \tilde{e}^-_R\tilde{e}^-_R$ shows a
strong dependence on $M_1$ \cite{Feng98,WER} this process offers also a 
possibility of its
measurement. The precision will, however,
depend on the mixing of the neutralino states.

The main focus of this paper is on the comparison of $e^-e^-$
scattering and $e^+e^-$ annihilation for the production of selectron
pairs.  It is organised as follows. In Sect.~2 we compare in the
framework of the MSSM the cross sections for the production of
selectron pairs $\tilde{e}_R\tilde{e}_R$, $\tilde{e}_L\tilde{e}_L$ and
$\tilde{e}_R \tilde{e}_L$ in $e^+e^-$ annihilation and $e^-e^-$
scattering. We take into account a general neutralino mixing and the
effects of initial state radiation (ISR) as well as beamstrahlung. We
emphasise the importance of beam polarization.  In Sect.~3 the
threshold behaviour of the cross sections for the three selectron
production channels is compared for the two linear collider modes with
regard to the measurement of the mass of the right and left selectron,
respectively. We demonstrate in Sect.~4 the possibility to establish
the partnership between chiral electrons and their scalar partners in
$e^+e^-$ annihilation with suitably polarized beams.  In Sect.~5 we
investigate the $M_1$ dependence of the production cross sections. In
Sect.~6 we discuss the implications of a possible mixing between left
and right selectrons.  The summary is given in Sect.~7. In the
Appendices we collect the formulas for the production cross sections
in the $e^+e^-$ and in the $e^-e^-$ mode for polarized beams as well
as for general selectron mixing.

\section{Impact of beam polarization on selectron pair production }
\label{sec:2}

We compare selectron production in $e^+ e^-$ annihilation via
$\gamma$-- and $Z^0$--exchange in the s--channel and
$\tilde{\chi}^0_i$--exchange ($i=1,\ldots,4$) in the crossed channel,
and in $e^- e^-$ scattering, where only $\tilde{\chi}^0_i$--exchange
contribute. We study the different selectron final states in
two MSSM scenarios.  In the following we use the GUT relation
$M_1=\frac{5}{3}\tan^2\Theta_W M_2$ for the gaugino mass parameters
except in Sect.~5.  In the first scenario (I) we take $M_2=152$~GeV,
$\mu=316$~GeV and $\tan\beta=3$.  Here the LSP $\tilde{\chi}^0_1$ is
bino--like and the lighter chargino $\tilde{\chi}^{\pm}_1$ is
wino--like. In the second scenario (II) with $M_2=$320~GeV,
$\mu=$150~GeV, $\tan\beta=3$ both the LSP and the lighter chargino are
higgsino--like. However, the LSP has a sizable bino--component.
The neutralino masses for both scenarios are given in
Table~\ref{tab_ez}. For the selectron masses we take
$m_{\tilde{e}_L}=$179.3~GeV and $m_{\tilde{e}_R}=$137.7~GeV.

\begin{table}
\begin{center}
\begin{tabular}{ccccccc}
\hline\noalign{\smallskip}
 scenario & $m_{\tilde{\chi}^\pm_1}$ & $m_{\tilde{\chi}^\pm_2}$ 
          & $m_{\tilde{\chi}^0_1}$ & $m_{\tilde{\chi}^0_2}$ 
          & $m_{\tilde{\chi}^0_3}$ & $m_{\tilde{\chi}^0_4}$ \\
\hline\noalign{\smallskip}
 (I) & 127.7 &    345.8 & 69.7 &    130.1 &     319.8 &     348.5 \\
 (II) &  126.4 &  349.1 & 102.1 &     154.2 &     183.3 &   349.8 \\
\hline
\end{tabular}
\end{center}
\caption{Chargino and neutralino masses  ([GeV]) for scenario (I) 
($M_2=152$~GeV, $\mu=316$~GeV, $\tan \beta=3$) and  scenario (II)
($M_2=320$~GeV, $\mu=150$~GeV, $\tan \beta=3$)
\label{tab_ez}}
\end{table}

\begin{table}
\begin{center}
\begin{tabular}{cccc}
\hline\noalign{\smallskip}
 scenario & $\nu_e {\tilde{\chi}^\pm_1}$ 
          & $e^- {\tilde{\chi}^0_1}$ & $e^- {\tilde{\chi}^0_2}$ \\
\hline\noalign{\smallskip}
 (I)  & 0.534 & 0.144 & 0.322  \\
 (II) & 0.986 & 0.011 & 0.003 \\
\hline
\end{tabular}
\end{center}
\caption{Branching ratios of $\selm{L}$ for both scenarios (I) and (II).
\label{tab_BR}}
\end{table}

The decays of the selectrons are already extensively studied
\cite{Bartl87}. The right
selectron couples mainly to the bino-component of the neutralino.
It decays therefore mainly in $\tilde{\chi}^0_i e$, where $i$ denotes the
neutralino with the largest
bino component. The left selectron decays
preferably into the wino-like chargino and a neutrino, followed by
a wino-like neutralino and an electron if all decays are
kinematically allowed.

\subsection{$e^+e^-$ Annihilation}\label{sec:2-1}

In $e^+e^-$ annihilation $\tilde{e}^+_R\tilde{e}^-_R$ and
$\tilde{e}^+_L\tilde{e}^-_L$ pairs are produced in a p--wave state with 
cross sections which rise at threshold as $\beta^3$, where
$\beta=2 p /\sqrt{s}$ is proportional to the selectron momentum:
\beqn
e^+_L e^-_R &\to& \tilde{e}^+_R \tilde{e}^-_R \label{eq_hans_1a},\\
e^+_R e^-_L &\to& \tilde{e}^+_L \tilde{e}^-_L \label{eq_hans_1b} \, .
\eeqn
The processes are mediated by $\gamma$/$Z$--exchange in the
s--channel and neutralino exchange in the t--channel \cite{Bartl87}.
In contrast to RR and LL selectron pairs the 
RL pairs are produced by electrons and positrons with the same chirality
\beqn
e^+_L e^-_L &\to& \tilde{e}^+_R \tilde{e}^-_L \label{eq_hans_2a},\\
e^+_R e^-_R &\to& \tilde{e}^+_L \tilde{e}^-_R \label{eq_hans_2b} \, .
\eeqn
They are produced in an s--wave state via t--channel exchange of neutralinos
with cross sections rising at threshold proportional to $\beta$.

\begin{figure*}
\setlength{\unitlength}{1mm}
\begin{center}
\begin{picture}(155,195)
\put(0,130){\mbox{\epsfig{figure=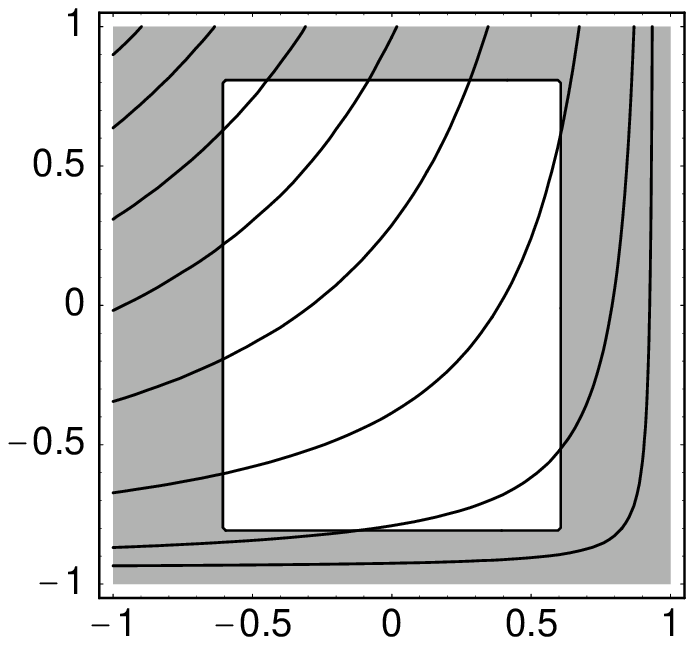,height=6.cm,width=6.cm}}}
\put(0,190){\makebox(0,0)[bl]{{{\bf a)}}}}
\put(5,189){\makebox(0,0)[bl]{{\bf $P_{e^-}$}}}
\put(17,189){\makebox(0,0)[bl]{$\sigma(e^+ e^- \to {\tilde e}^-_R 
{\tilde e}^+_R)$ [fb]}}
\put(62,129){\makebox(0,0)[br]{{\bf $P_{e^+}$}}}
\put(57,139){\makebox(0,0)[br]{{\small 50}}}
\put(48,143){\makebox(0,0)[br]{{\small 100}}}
\put(41,150){\makebox(0,0)[br]{{\small 250}}}
\put(34,160){\makebox(0,0)[br]{{\small 500}}}
\put(28,166){\makebox(0,0)[br]{{\small 750}}}
\put(27,175){\makebox(0,0)[br]{{\small 1000}}}
\put(17,175){\makebox(0,0)[br]{{\small 1250}}}
\put(16,182){\makebox(0,0)[br]{{\small 1450}}}
\put(75,130){\mbox{\epsfig{figure=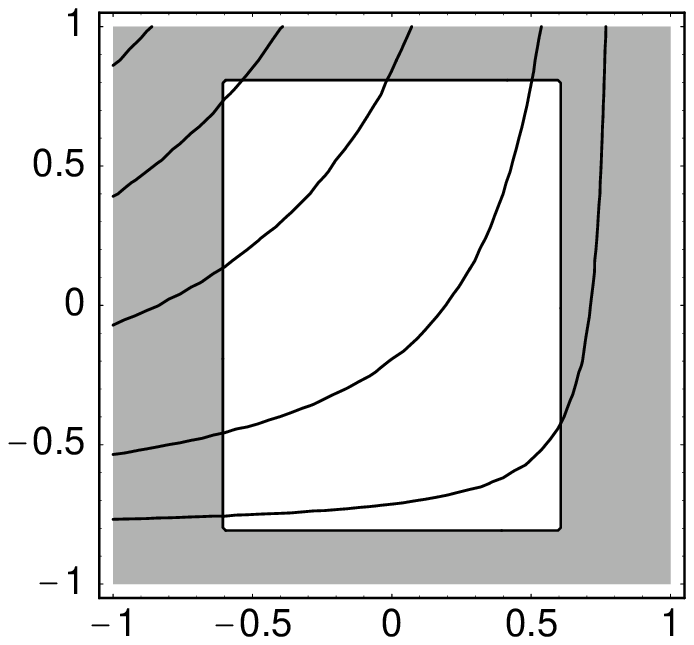,height=6.cm,width=6.cm}}}
\put(91,189){\makebox(0,0)[bl]{$\sigma(e^+ e^- \to {\tilde e}^-_R 
{\tilde e}^+_R)$ [fb]}}
\put(75,190){\makebox(0,0)[bl]{{{\bf b)}}}}
\put(80,189){\makebox(0,0)[bl]{{\bf $P_{e^-}$}}}
\put(137,129){\makebox(0,0)[br]{{\bf $P_{e^+}$}}}
\put(122,144){\makebox(0,0)[br]{{50}}}
\put(115,155){\makebox(0,0)[br]{{100}}}
\put(107,167){\makebox(0,0)[br]{{200}}}
\put(101,177){\makebox(0,0)[br]{{ 300}}}
\put(93,182){\makebox(0,0)[br]{{ 400}}}
\put(0,65){\mbox{\epsfig{figure=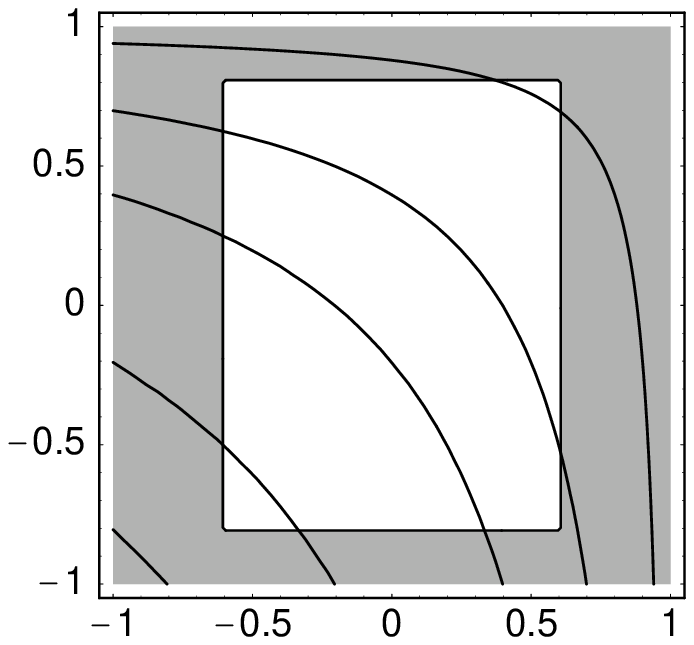,height=6.cm,width=6.cm}}}
\put(0,125){\makebox(0,0)[bl]{{{\bf c)}}}}
\put(5,124){\makebox(0,0)[bl]{{\bf $P_{e^-}$}}}
\put(17,124){\makebox(0,0)[bl]{$\sigma(e^+ e^- \to {\tilde e}^-_L 
{\tilde e}^+_R)$ [fb]}}
\put(63,64){\makebox(0,0)[br]{{\bf $P_{e^+}$}}}
\put(18,75){\makebox(0,0)[br]{{300}}}
\put(28,83){\makebox(0,0)[br]{{200}}}
\put(38,95){\makebox(0,0)[br]{{100}}}
\put(44,103){\makebox(0,0)[br]{{ 50}}}
\put(54,115){\makebox(0,0)[br]{{ 10}}}
\put(75,65){\mbox{\epsfig{figure=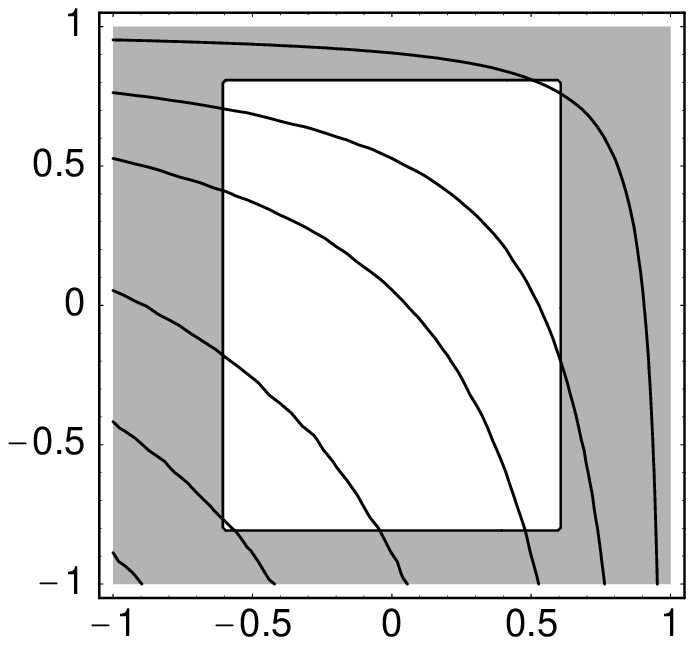,height=6.cm,width=6.cm}}}
\put(80,125){\makebox(0,0)[bl]{{{\bf d)}}}}
\put(83,124){\makebox(0,0)[bl]{{\bf $P_{e^-}$}}}
\put(92,124){\makebox(0,0)[bl]{$\sigma(e^+ e^- \to {\tilde e}^-_L 
{\tilde e}^+_R)$ [fb]}}
\put(137,64){\makebox(0,0)[br]{{\bf $P_{e^+}$}}}
\put(92,74){\makebox(0,0)[br]{{400}}}
\put(101,80){\makebox(0,0)[br]{{300}}}
\put(107,88){\makebox(0,0)[br]{{200}}}
\put(115,99){\makebox(0,0)[br]{{100}}}
\put(120,107){\makebox(0,0)[br]{{ 50}}}
\put(129,117){\makebox(0,0)[br]{{ 10}}}
\put(0,0){\mbox{\epsfig{figure=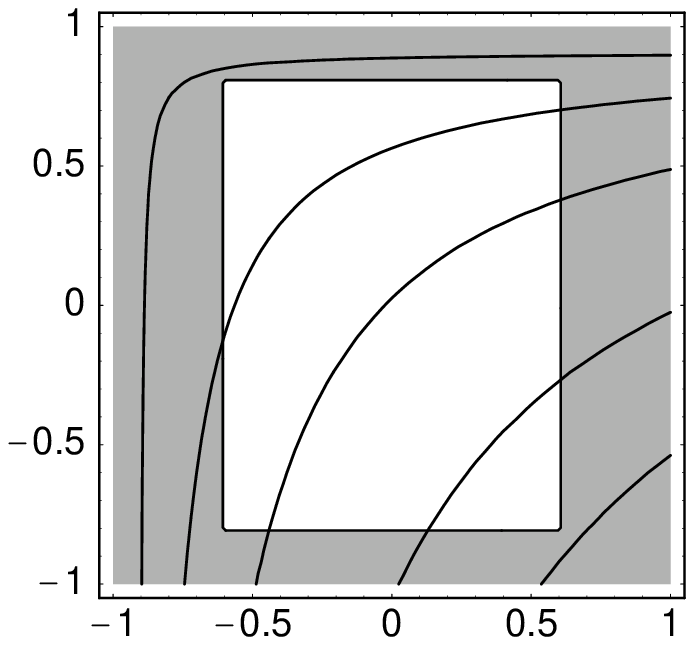,height=6.cm,width=6.cm}}}
\put(0,60){\makebox(0,0)[bl]{{{\bf e)}}}}
\put(5,59){\makebox(0,0)[bl]{{\bf $P_{e^-}$}}}
\put(17,59){\makebox(0,0)[bl]{$\sigma(e^+ e^- \to {\tilde e}^-_L 
{\tilde e}^+_L)$ [fb]}}
\put(62,-1){\makebox(0,0)[br]{{\bf $P_{e^+}$}}}
\put(55,17){\makebox(0,0)[br]{{300}}}
\put(43,22){\makebox(0,0)[br]{{200}}}
\put(34,34){\makebox(0,0)[br]{{100}}}
\put(25,43){\makebox(0,0)[br]{{ 50}}}
\put(15,52){\makebox(0,0)[br]{{ 20}}}
\put(75,0){\mbox{\epsfig{figure=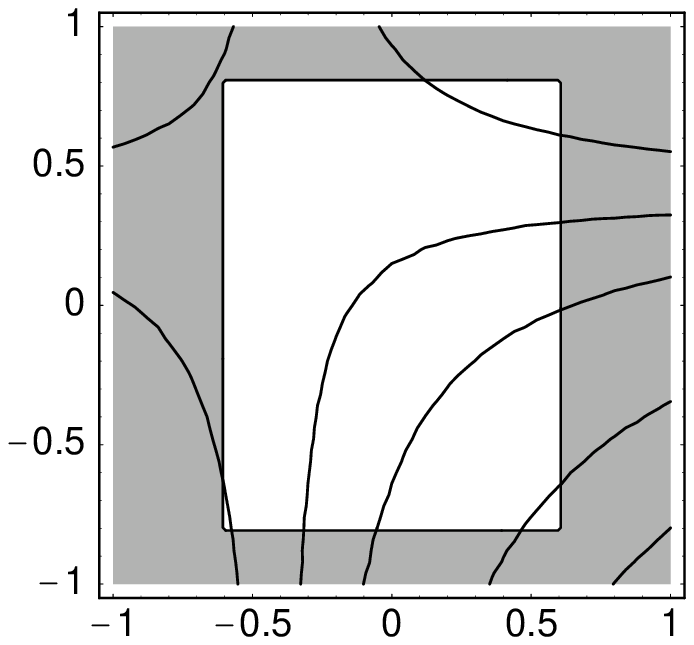,height=6.cm,width=6.cm}}}
\put(75,60){\makebox(0,0)[bl]{{{\bf f)}}}}
\put(80,59){\makebox(0,0)[bl]{{\bf $P_{e^-}$}}}
\put(92,59){\makebox(0,0)[bl]{$\sigma(e^+ e^- \to {\tilde e}^-_L 
{\tilde e}^+_L)$ [fb]}}
\put(137,-1){\makebox(0,0)[br]{{\bf $P_{e^+}$}}}
\put(130,11){\makebox(0,0)[br]{{40}}}
\put(122,17){\makebox(0,0)[br]{{30}}}
\put(114,27){\makebox(0,0)[br]{{20}}}
\put(107,35){\makebox(0,0)[br]{{ 15}}}
\put(129,48){\makebox(0,0)[br]{{ 10}}}
\put(91,51){\makebox(0,0)[br]{{ 15}}}
\put(91,33){\makebox(0,0)[br]{{ 10}}}
\end{picture}
\end{center}
\caption[]{Contour lines for the cross sections in fb for
  {\bf a)} and {\bf b)} $\sigma(e^+ e^- \to {\tilde e}^-_R {\tilde
    e}^+_R)$, {\bf c)} and {\bf d)} $\sigma(e^+ e^- \to {\tilde e}^-_L
  {\tilde e}^+_R)$, {\bf e)} and {\bf f)} $\sigma(e^+ e^- \to {\tilde
    e}^-_L {\tilde e}^+_L)$ as a function of electron polarization
  $P_{e^-}$ and positron polarization $P_{e^+}$ for $\sqrt{s} =
  500$~GeV, $m_{\tilde{e}_R}=137.7$ GeV, $m_{\tilde{e}_L}=179.3$ GeV,
  $\tan\beta=3$; in {\bf a)}, {\bf c)} and {\bf e)} $M_2=152$~GeV,
  $\mu=316$~GeV and in {\bf b)}, {\bf d)} and {\bf f)} $M_2=320$~GeV,
  $\mu=150$~GeV. ISR corrections and beamstrahlung are included. The
  white area shows the region which can be realized
  within the TESLA design with $|P_{e^-}|=0.8$, $|P_{e^+}|=0.6$.  }
\label{fig:PolepemH}
\end{figure*}

We show for $\sqrt{s}=500$~GeV the dependence on
beam polarization of the cross sections for the different production channels
for scenario (I) in \fig{fig:PolepemH}a, c, and e and for
scenario (II) in \fig{fig:PolepemH}b, d, and f. 
The effects of ISR and beamstrahlung have been included. 
For ISR we use the structure function prescription 
\cite{Skrzypek:1991qs} and for 
beamstrahlung we generate the
spectrum from the approximate integral equation given in \cite{myLC}.
In the white areas
which corresponds to the proposed TESLA design \cite{TDR} the cross
sections in scenario (I) for $e^+ e^-\to \tilde{e}^+_R \tilde{e}^-_R$ and
$e^+ e^-\to \tilde{e}^+_{R} \tilde{e}^-_{L}$ vary by about a factor 20 and that
for $e^+ e^-\to \tilde{e}^+_L \tilde{e}^-_L$ by about a factor 10. For right
polarized electrons ($P_{e^-}=0.8$) and left polarized positrons 
($P_{e^+}=-0.6$) one obtains the highest cross section of 1340~fb for the
production of $\tilde{e}^+_R \tilde{e}^-_R$ pairs. Those for   
the other pairs $\tilde{e}^+_{R} \tilde{e}^-_{L}$ and 
$\tilde{e}^+_L \tilde{e}^-_L$ lead to rates less than 30~fb.
In the scenario (II) the rates for $\tilde{e}^+_R \tilde{e}^-_R$ 
and $\tilde{e}^+_L \tilde{e}^-_L$ are significantly smaller
than in scenario (I) because in this case the interference between the s-- 
and the t--channel is more destructive.
The rates for
$\tilde{e}^+_R \tilde{e}^-_L$ increase. 
The polarization dependence of 
$\sigma(e^+ e^- \to \tilde{e}^+_L \tilde{e}^-_R)$
is obtained by reversing the sign of $P_{e^-}$ and $P_{e^+}$ 
in \fig{fig:PolepemH}c and d,
as a consequence of the pure t--channel exchange. 
\begin{figure*}
\setlength{\unitlength}{1mm}
\begin{center}
\begin{picture}(155,195)
\put(0,130){\mbox{\epsfig{figure=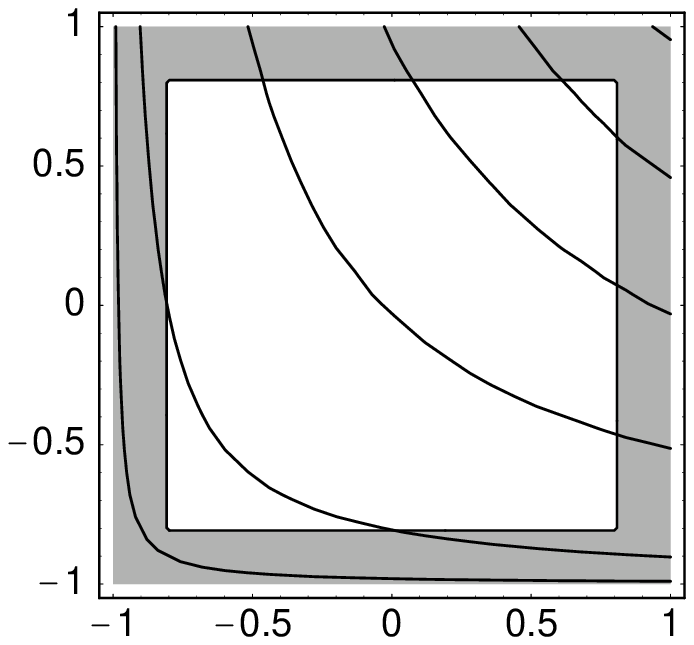,height=6.cm,width=6.cm}}}
\put(0,190){\makebox(0,0)[bl]{{{\bf a)}}}}
\put(5,189){\makebox(0,0)[bl]{{\bf $P_{e_1}$}}}
\put(17,189){\makebox(0,0)[bl]{$\sigma(e^- e^- \to {\tilde e}^-_R
{\tilde e}^-_R)$ [pb]}}
\put(62,129){\makebox(0,0)[br]{{\bf $P_{e_2}$}}}
\put(18,145){\makebox(0,0)[br]{{0.01}}}
\put(27,148){\makebox(0,0)[br]{{0.1}}}
\put(37,157){\makebox(0,0)[br]{{0.5}}}
\put(44,168){\makebox(0,0)[br]{{1.0}}}
\put(52,177){\makebox(0,0)[br]{{1.5}}}
\put(58,183){\makebox(0,0)[br]{{2}}}
\put(75,130){\mbox{\epsfig{figure=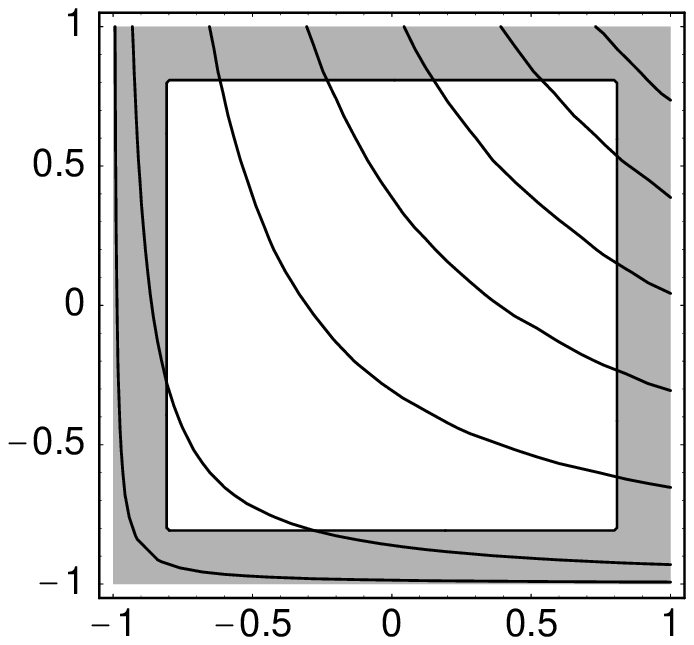,height=6.cm,width=6.cm}}}
\put(91,189){\makebox(0,0)[bl]{$\sigma(e^- e^- \to {\tilde e}^-_R
{\tilde e}^-_R)$ [pb]}}
\put(75,190){\makebox(0,0)[bl]{{{\bf b)}}}}
\put(80,189){\makebox(0,0)[bl]{{\bf $P_{e_1}$}}}
\put(137,129){\makebox(0,0)[br]{{\bf $P_{e_2}$}}}
\put(96,139){\makebox(0,0)[br]{{0.01}}}
\put(100,146){\makebox(0,0)[br]{{0.1}}}
\put(107,154){\makebox(0,0)[br]{{0.5}}}
\put(114,163){\makebox(0,0)[br]{{1}}}
\put(120,170){\makebox(0,0)[br]{{1.5}}}
\put(126,175){\makebox(0,0)[br]{{2}}}
\put(127,183){\makebox(0,0)[br]{{2.5}}}
\put(0,65){\mbox{\epsfig{figure=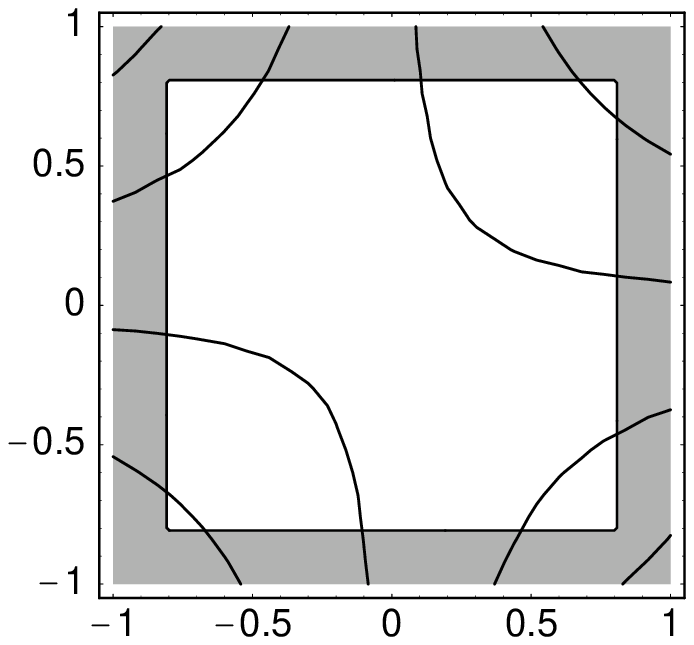,height=6.cm,width=6.cm}}}
\put(0,125){\makebox(0,0)[bl]{{{\bf c)}}}}
\put(5,124){\makebox(0,0)[bl]{{\bf $P_{e_1}$}}}
\put(17,124){\makebox(0,0)[bl]{$\sigma(e^- e^- \to {\tilde e}^-_R
{\tilde e}^-_L)$ [pb]}}
\put(63,64){\makebox(0,0)[br]{{\bf $P_{e_2}$}}}
\put(21,79){\makebox(0,0)[br]{{0.1}}}
\put(31,92){\makebox(0,0)[br]{{0.2}}}
\put(18,118){\makebox(0,0)[br]{{0.4}}}
\put(24,110){\makebox(0,0)[br]{{0.3}}}
\put(45,106){\makebox(0,0)[br]{{ 0.2}}}
\put(52,113){\makebox(0,0)[br]{{ 0.1}}}
\put(48,82){\makebox(0,0)[br]{{0.3}}}
\put(56,75){\makebox(0,0)[br]{{0.4}}}
\put(75,65){\mbox{\epsfig{figure=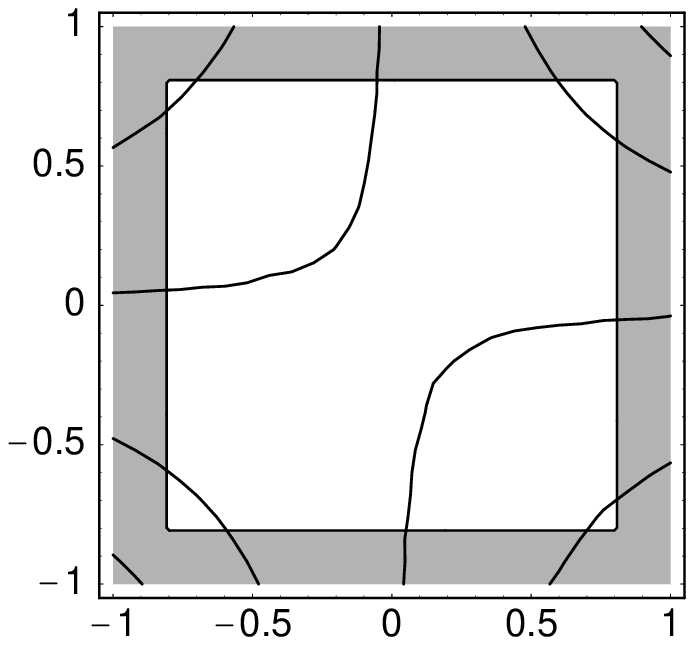,height=6.cm,width=6.cm}}}
\put(80,125){\makebox(0,0)[bl]{{{\bf d)}}}}
\put(83,124){\makebox(0,0)[bl]{{\bf $P_{e_1}$}}}
\put(92,124){\makebox(0,0)[bl]{$\sigma(e^- e^- \to {\tilde e}^-_R
{\tilde e}^-_L)$ [pb]}}
\put(137,64){\makebox(0,0)[br]{{\bf $P_{e_2}$}}}
\put(90.5,74){\makebox(0,0)[br]{{0.1}}}
\put(97,80){\makebox(0,0)[br]{{0.2}}}
\put(110,100){\makebox(0,0)[br]{{0.3}}}
\put(95,113){\makebox(0,0)[br]{{0.4}}}
\put(132,117){\makebox(0,0)[br]{{ 0.1}}}
\put(126,111){\makebox(0,0)[br]{{ 0.2}}}
\put(114,93){\makebox(0,0)[br]{{0.3}}}
\put(128,80){\makebox(0,0)[br]{{0.4}}}
\put(0,0){\mbox{\epsfig{figure=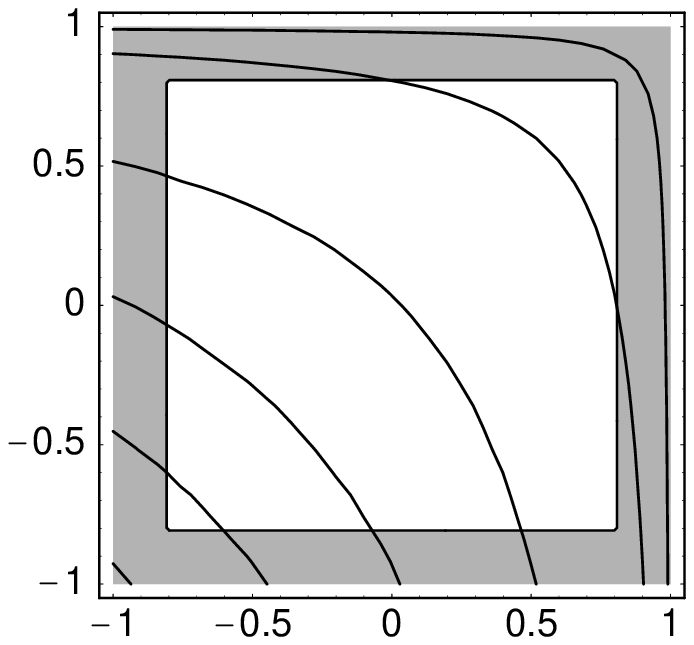,height=6.cm,width=6.cm}}}
\put(0,60){\makebox(0,0)[bl]{{{\bf e)}}}}
\put(5,59){\makebox(0,0)[bl]{{\bf $P_{e_1}$}}}
\put(17,59){\makebox(0,0)[bl]{$\sigma(e^- e^- \to {\tilde e}^-_L
{\tilde e}^-_L)$ [pb]}}
\put(62,-1){\makebox(0,0)[br]{{\bf $P_{e_2}$}}}
\put(12,9){\makebox(0,0)[br]{{2}}}
\put(22,15){\makebox(0,0)[br]{{1.5}}}
\put(30,22){\makebox(0,0)[br]{{1.0}}}
\put(41,32){\makebox(0,0)[br]{{0.5}}}
\put(49,40){\makebox(0,0)[br]{{0.1}}}
\put(47,52.5){\makebox(0,0)[br]{{0.01}}}
\put(75,0){\mbox{\epsfig{figure=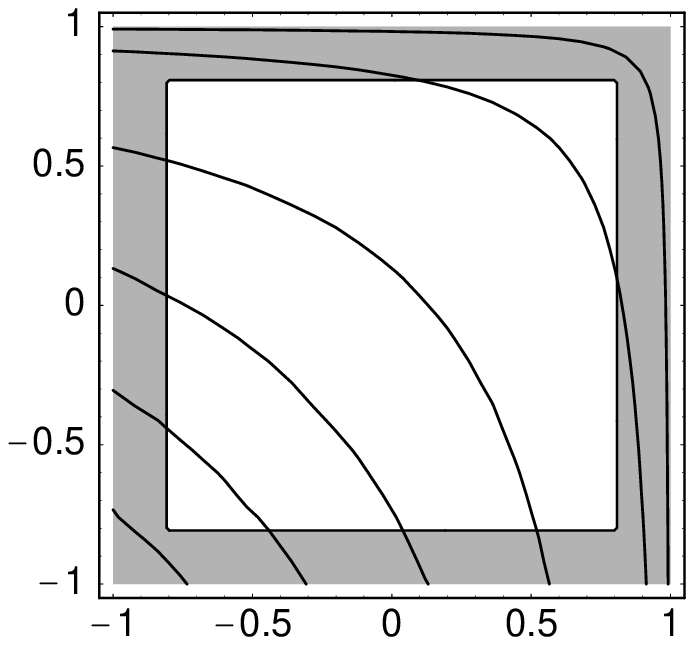,height=6.cm,width=6.cm}}}
\put(75,60){\makebox(0,0)[bl]{{{\bf f)}}}}
\put(80,59){\makebox(0,0)[bl]{{\bf $P_{e_1}$}}}
\put(92,59){\makebox(0,0)[bl]{$\sigma(e^- e^- \to {\tilde e}^-_L
{\tilde e}^-_L)$ [pb]}}
\put(137,-1){\makebox(0,0)[br]{{\bf $P_{e_2}$}}}
\put(93,9){\makebox(0,0)[br]{{2}}}
\put(100,16){\makebox(0,0)[br]{{1.5}}}
\put(106,23){\makebox(0,0)[br]{{1}}}
\put(116,33){\makebox(0,0)[br]{{0.5}}}
\put(124,43){\makebox(0,0)[br]{{ 0.1}}}
\put(125,52.5){\makebox(0,0)[br]{{ 0.01}}}
\end{picture}
\end{center}
\caption[]{Contour lines for the cross sections in pb for
  {\bf a)} and {\bf b)} $\sigma(e^- e^- \to {\tilde e}^-_R {\tilde
    e}^-_R)$, {\bf c)} and {\bf d)} $\sigma(e^- e^- \to {\tilde e}^-_R
  {\tilde e}^-_L)$, {\bf e)} and {\bf f)} $\sigma(e^- e^- \to {\tilde
    e}^-_L {\tilde e}^-_L)$ as a function of electron polarizations
  $P_{e_{1/2}}$ for $\sqrt{s} =
  500$~GeV, $m_{\tilde{e}_R}=137.7$ GeV, $m_{\tilde{e}_L}=179.3$ GeV,
  $\tan\beta=3$; in {\bf a)}, {\bf c)} and {\bf e)} $M_2=152$~GeV,
  $\mu=316$~GeV and in {\bf b)}, {\bf d)} and {\bf f)} $M_2=320$~GeV,
  $\mu=150$~GeV. ISR corrections and beamstrahlung are included. The
  white area shows the region which can be realized
  within the TESLA design with $|P_{e_{1,2}}|=0.8$.  }
\label{fig:e-e-pol}
\end{figure*}

\subsection{$e^-e^-$ Scattering}
Right and left selectrons are produced by right-- and left--handed electrons
since only t-- and u--channel exchange of neutralinos contribute:
\begin{eqnarray}
e^-_{R} e^-_{R} &\to& \tilde{e}^-_{R} \tilde{e}^-_{R}, \label{eq_emem_1a} \\
e^-_{L} e^-_{L} &\to& \tilde{e}^-_{L} \tilde{e}^-_{L}, \label{eq_emem_1c} \\
e^-_R e^-_L &\to& \tilde{e}^-_R \tilde{e}^-_L. \label{eq_emem_1b}
\end{eqnarray}
In contrast to $e^+ e^-$ annihilation RR and LL pairs are produced in a
s--wave state with the cross sections at threshold rising as $\beta$, whereas
RL pairs  are produced in a p--wave state with a $\beta^3$
behaviour of the threshold cross section.

The polarization dependence of the cross sections
is depicted in \fig{fig:e-e-pol}a, c, e for scenario (I)
and   in \fig{fig:e-e-pol}b, d, f for scenario (II)
for $\sqrt{s}=500$~GeV. Here we have taken into account the effects of
ISR and beamstrahlung. For ISR we use the structure function prescription 
\cite{Skrzypek:1991qs} and for 
beamstrahlung we use the energy distribution given by Eq.~(6) in \cite{myLC}
evaluated with the approximation P2. For the luminosity enhancement
parameter we use the analytic approximation given by Eq.~(14) in 
\cite{Thompson}.

Quite generally the cross sections are larger than for
selectron production by $e^+ e^-$ annihilation. For the highest polarization
in the TESLA design, $P_{e_{1,2}}=\pm 0.8$ for the electron beams, one
obtains for the production of $\tilde{e}^-_R \tilde{e}^-_R$ and 
$\tilde{e}^-_L \tilde{e}^-_L$  cross sections of about 2.5 pb for
$P_{e_{1,2}}=0.8$ and $P_{e_{1,2}}=-0.8$, respectively. For the
production of left--right selectron pairs a maximal cross section 
of  360 fb in scenario (I) and 400 fb in scenario (II) is reached
for $P_{e_1}=0.8$, $P_{e_2}=-0.8$. 

According to Eqs.~(\ref{eq_emem_1a}) -- (\ref{eq_emem_1b}) 
the cross section for RR (LL) pairs vanishes if one of the electron beams is
completely left--handed (right--handed) polarized and that for RL pairs 
vanishes if both beams are completely polarized with the same sign.
 
\section{Threshold behaviour}
\label{sec:3}
The precision on the determination of the masses of $\tilde{e}_R$ and
$\tilde{e}_L$ in $e^+e^-$ annihilation and $e^- e^-$ scattering,
respectively, depends on both the threshold behaviour of the cross
sections for the different production channels and on  a suitable choice
of beam polarizations. 
As we are mainly interested in the comparison of
the collider modes we take into account statistical errors only
disregarding effects of the finite widths of the selectrons, which 
will be studied in future investigations \cite{Ayres}, and of beam energy
spread which would need Monte Carlo simulations.
The comparison is done for scenario (I) of
Table~\ref{tab_ez}. We comment on scenario (II) at the end of this section.
The determination of  selectron masses in the continuum is discussed in
\cite{Dima:2001jr}.

\subsection{Measurement of $m_{\tilde{e}_R}$}\label{sec:3-1}

In scenario (I) the right selectron decays nearly always into an
electron and the LSP. Therefore, the signature of $\tilde{e}_R$ pair
production in $e^+e^-$ annihilation ($e^-e^-$ scattering) is an
$e^+e^-$ ($e^-e^-$) pair and missing energy.  In the following we
assume an effective luminosity of 10~fb$^{-1}$ in case of $e^+ e^-$
and an effective luminosity of 1~fb$^{-1}$ in case of $e^- e^-$ at each 
scanning step.

\subsubsection{$e^+e^-$ Annihilation}\label{sec:3-1-1}

The favourable polarization is $P_{e^-}=+0.8$ and  
$P_{e^+}=-0.6$.
In this case one obtains both the highest cross sections for
$e^+ e^-\to \tilde{e}_R^+ \tilde{e}_R^-$ and maximal suppression 
of the SM background from $W$ and $Z^0$ pair production 
and from single 
$W$ production. Also for this configuration 
of beam polarizations the SUSY background processes
$e^+ e^-\to \tilde{\chi}^0_1 \tilde{\chi}^0_2
        \to \tilde{\chi}^0_1 \tilde{\chi}^0_1 e^+ e^-$ 
and $e^+ e^- \to \tilde{\chi}^0_2 \tilde{\chi}^0_2 \to 
\tilde{\chi}^0_1 \tilde{\chi}^0_1 \nu_e \bar{\nu}_e e^+ e^-$ 
are reduced.

\begin{figure}
\setlength{\unitlength}{1mm}
\begin{center}
\begin{picture}(83,70)
\put(0,-2){\mbox{\epsfig{figure=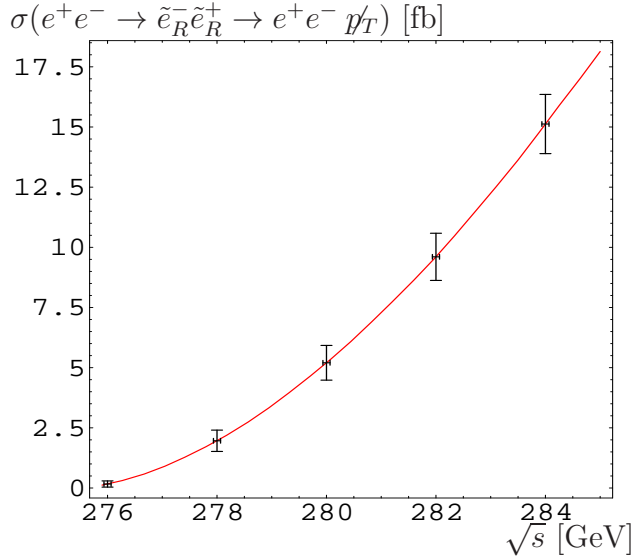,height=7cm,width=8cm}}}
\put(0,66){\makebox(0,0)[bl]{{$\sigma(e^+e^-
  \to \tilde{e}^-_{R}\tilde{e}^+_{R} \to  e^+e^-\not\!\!p_T$)~[fb]}}}
\put(83,-3){\makebox(0,0)[br]{{$\sqrt{s}$~[GeV]}}}
\end{picture}
\end{center}
\caption[]{
  Threshold behaviour of the process
  $e^+e^- \to \tilde{e}^-_{R}\tilde{e}^+_{R} \to 
  e^+e^-\not\!\!p_T$ for $P_{e^-}=+0.8$ and
  $P_{e^+}=-0.6$, $M_2 = 152$~GeV, $\mu=316$~GeV and $\tanb=3$.
  ISR corrections and beamstrahlung are included. The
  error bars show the statistical error for $\mathcal{L}=10$ fb$^{-1}$. 
  }
\label{the+e-1}
\end{figure}
\begin{figure}
\setlength{\unitlength}{1mm}
\begin{center}
\begin{picture}(83,70)
\put(0,-1){\mbox{\epsfig{
  figure=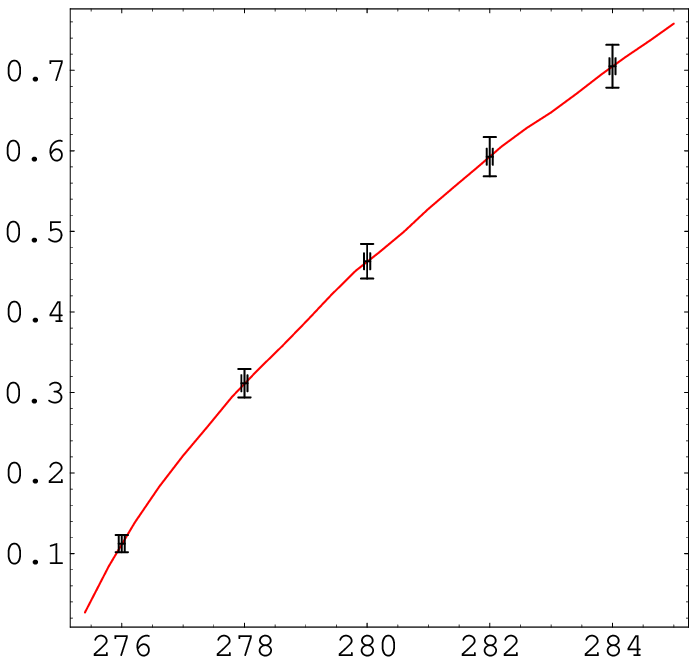,height=6.7cm,width=8cm}}}
\put(0,66){\makebox(0,0)[bl]{{$\sigma(e^-e^- \to \tilde{e}^-_{R}\tilde{e}^-_{R}
                           \to e^-e^-\not\!\!p_T$)~[pb]}}}
\put(83,-3){\makebox(0,0)[br]{{$\sqrt{s}$~[GeV]}}}
\end{picture}
\end{center}
\caption[]{
  Threshold behaviour of the process $e^-e^- \to
  \tilde{e}^-_{R}\tilde{e}^-_{R} \to e^-e^-\not\!\!p_T$ for
  polarizations: $P_{e_1}=0.8$ and $P_{e_2}=0.8$.
 ISR corrections and beamstrahlung are included. The error bars show the
  statistical error for $\mathcal{L}=1$ fb$^{-1}$.}
\label{the-e-1}
\end{figure}
\fig{the+e-1} shows for these polarizations the cross sections
near threshold for $e^+e^-\to \tilde{e}_R^+ \tilde{e}_R^- \to e^+ e^-
\tilde{\chi}^0_1 \tilde{\chi}^0_1$ for $m_{\tilde{e}_R}=137.7$~GeV,
and the $1\sigma$ error bars for an effective luminosity
of 
${\cal L}=10$fb$^{-1}$ for each point. The horizontal error bars
indicate the expected errors on the selectron mass.
 Provided the neutralino mass parameters are precisely known the measurement of
the cross section at five points results in an error of $\Delta
m_{\tilde{e}_R}=65$~MeV.
Since the statistical error is
comparable to the influence of the finite width of the selectrons and
of the Coulomb rescattering \cite{Ayres1,Ayres} 
these effects have to be
taken into account for a precise determination of $m_{\tilde{e}_R}$.
Note that this precise determination allows a precise determination
of the underlying SUSY parameters. This in turn allows for a precise 
determination of parameters at high energy scales \cite{Blair:2000gy}.

\subsubsection{$e^-e^-$ Scattering}\label{sec:3-1-2}

In $e^-e^-$ scattering the favourable beam polarization leading to
the largest cross sections is $P_{e_1}=P_{e_2}=0.8$.
In this case the most important backgrounds  
$e^- e^-\to W^- e^- \nu_e$, $W^- W^- \nu_e \nu_e$ from SM processes and
$e^- e^-\to \tilde{\chi}^-_1 \tilde{\chi}^0_1 e^- \nu_e$ 
from SUSY are suppressed. For this purpose
simultaneous polarization of both beams is
important.  

In spite of the lower luminosity in $e^- e^-$ scattering compared to
$e^+e^-$ annihilation the steep $\beta$ rise of the cross section at
threshold allows the determination of the mass of $\tilde{e}_R$ with a
slightly higher precision, $m_{\tilde{e}_R}=137.7\pm 0.05$~GeV, which
is demonstrated in \fig{the-e-1} for a beam polarizations
$P_{e_1}=P_{e_2}=0.8$.  The error bars correspond to an effective
luminosity ${\cal L}=1$~fb$^{-1}$ for each point.  The
statistical error is as in the case of an $e^+ e^-$ collider
comparable to the influence of the finite width of the selectrons and
of the Coulomb rescattering which have to be taken into account for a
precise determination of $m_{\tilde{e}_R}$ \cite{Ayres1,Ayres}.

\subsection{Measurement of $m_{\tilde{e}_L}$}\label{sec:3-2}

\subsubsection{$e^+e^-$ Annihilation}\label{sec:3-2-1}

In principle the left selectron mass can be determined by a threshold
scan of either $\tilde{e}_R^{\pm}\tilde{e}_L^{\mp}$ production or 
$\tilde{e}_L^+\tilde{e}_L^-$ production. If
$m_{\tilde{e}_R}$ is already known the production of a pair of
left--right selectrons is more suitable due to the threshold behaviour
with the steeper $\beta$ rise. For $e^+ e^-\to
\tilde{e}_R^+ \tilde{e}_L^-$ ($\tilde{e}_L^+ \tilde{e}_R^-$) the
polarization $P_{e^-}=-0.8$, $P_{e^+}=-0.6$ ($P_{e^-}=0.8$,
$P_{e^+}=0.6$) leads to the largest cross sections.

In our scenario $\tilde{e}_R^-$ decays into an electron
and the LSP, whereas for the left selectron the following decays are
allowed: $\tilde{e}_L^- \to e \tilde \chi^0_1$, $\tilde{e}_L^- \to e
\tilde \chi^0_2$ and $\tilde{e}_L^- \to \nu_e \tilde \chi^-_1$. The decays 
of the
chargino $\tilde \chi^-_1$ as well as the neutralino $\tilde \chi^0_2$
decay lead to four-particle final states.

The final state $e^+e^- p_T \hspace {-3.5mm}/$ \, is not suitable for
the identification of the $\tilde{e}_R \tilde{e}_L$ pair since the
background from $\tilde{e}^+_R \tilde{e}^-_R$ production is
significantly larger than the signal $e^+ e^- \to \tilde{e}^+_R
\tilde{e}^-_L + \tilde{e}^-_R \tilde{e}^+_L \to e^+ e^- \tilde
\chi^0_1 \tilde \chi^0_1 + e^+ e^- \tilde \chi^0_1 \tilde \chi^0_1 \nu
\nu$. Even for $P_{e^-}=-0.8$, $P_{e^+}=-0.6$, where the signal is largest,
the ratio of the cross sections RR:RL is 5:1.
The final state should contain an $e^+e^-$ pair for the
identification of the flavour of the produced sleptons. Therefore,
we consider the decay $\tilde{e}^\pm_L \to e^\pm \tilde \chi^0_2 \to
e^{\pm} j j p_T \hspace{-3.5mm} /$ \,, where $j$ denotes a jet and we
assume a $BR(\tilde{\chi}^0_2\to j j \tilde{\chi}^0_1)= 0.35$.  The
main SM background stems from triple gauge boson production $W^+ W^-
Z$, $W^+ W^- \gamma^*$ and $ZZZ$.  The cross section for these
processes leading to the final state $e^+ e^- j j p_T \hspace{-3.5mm}
/$ \, is below $1$ fb \cite{Gunion}.  It is further suppressed by our
choice of beam polarization.  The main SUSY background from $e^+ e^-
\to \tilde \chi^0_2 \tilde \chi^0_2 \to e^+ e^- j j p_T
\hspace{-3.5mm} /$ \, has in general a different event topology compared
to our signal and can also be suppressed by an appropriate choice of
beam polarization.

\begin{figure}
\setlength{\unitlength}{1mm}
\begin{center}
\begin{picture}(83,70)
\put(0,0){\mbox{\epsfig{
                  figure=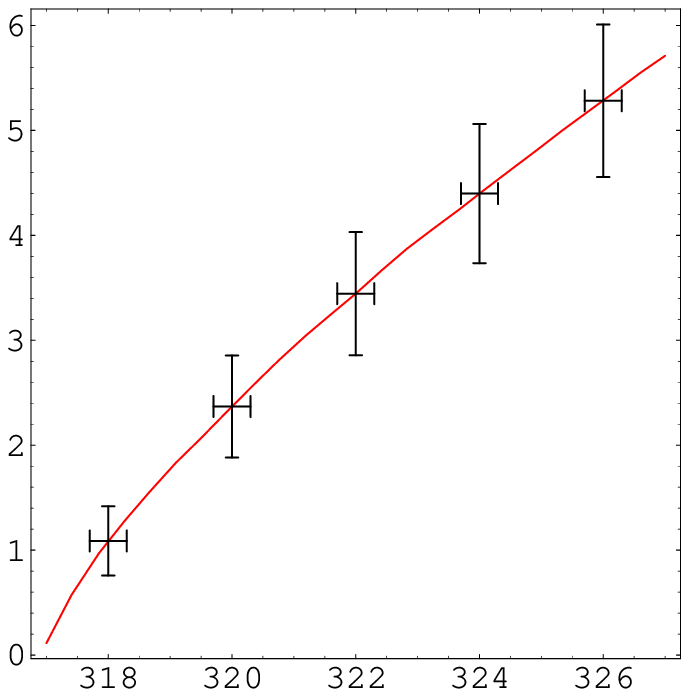,height=6.5cm,width=8.cm}}}
\put(0,66){\makebox(0,0)[bl]{{$\sigma(e^+e^- 
   \to \tilde{e}^+_{L}\tilde{e}^-_{R} +\tilde{e}^-_{L}\tilde{e}^+_{R}
   \to e^- e^+ j j \not\!\!p_T)$~[fb]}}}
\put(83,-3){\makebox(0,0)[br]{{$\sqrt{s}$~[GeV]}}}
\end{picture}
\end{center}
\caption[]{Threshold behaviour of the processes 
  $e^+e^- \to \tilde{e}^+_{L}\tilde{e}^-_{R}
  +\tilde{e}^-_{L}\tilde{e}^+_{R} \to e^- e^+ j j \not\!\!p_T$ for
  $P_{e^-}=0.8$ and $P_{e^+}=0.6$, $M_2 = 152$~GeV, $\mu=316$~GeV and
  $\tanb=3$.  ISR corrections and beamstrahlung are included.  The
  error bars show the statistical error for $\mathcal{L}=10$ fb$^{-1}$.
  }
\label{the+e-2}
\end{figure}

We show in \fig{the+e-2} the threshold for
$e^+ e^- \to \tilde{e}_R^-\tilde{e}_L^+ + \tilde{e}_R^+\tilde{e}_L^-
\to  e^+ e^- j j  p_T \hspace{-3.5mm} /$ \,
 for scenario (I) and $P_{e^-}=0.8$, $P_{e^+}=0.6$ so that
 the signal is maximal and the
SUSY background is minimal. 
Assuming 5 data points with an integrated 
luminosity of 10~fb$^{-1}$ for each
we find a statistical error of  $\Delta \msel{\pm}{L} = 300$~MeV.

The cross section for $e^+ e^- \to \tilde{e}_L^-\tilde{e}_L^+ \to e^+
e^- j j j j p_T \hspace{-3.5mm} /$ \, is below 0.2 fb near 
threshold. This small cross section is caused by the $\beta^3$
dependence at threshold and  by the small
branching ratio of $\tilde{e}_L$ into $e j j p_T \hspace{-3.5mm} /$ \, in our
scenario. This
implies a rather large statistical error resulting in $\Delta
\msel{\pm}{L} = 1.2$~GeV which is a factor 4 worse compared to the
measurement at the $\tilde{e}_L^\pm\tilde{e}_R^\mp$ threshold.

\subsubsection{$e^-e^-$ Scattering}\label{sec:3-2-2}

\begin{figure}
\setlength{\unitlength}{1mm}
\begin{center}
\begin{picture}(83,70)
\put(0,-1){\mbox{\epsfig{
                  figure=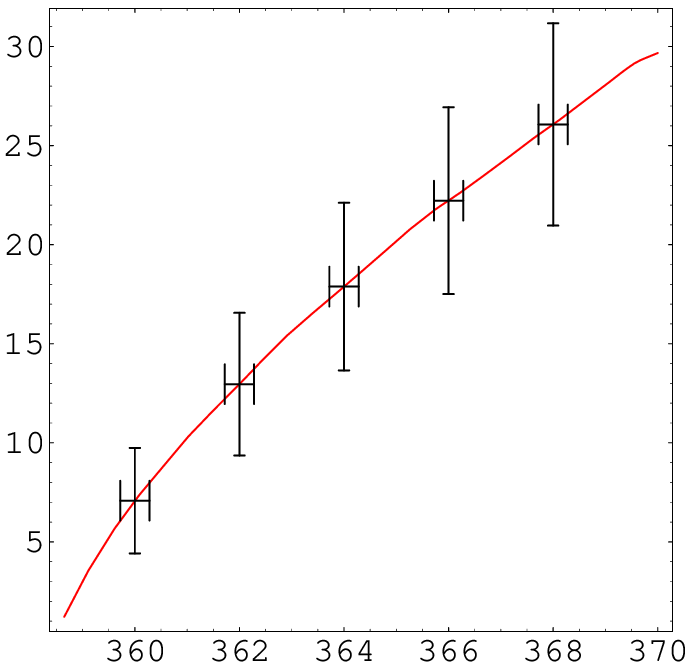,height=6.7cm,width=8.cm}}}
\put(0,66){\makebox(0,0)[bl]{{$\sigma(e^-e^- 
\to \tilde{e}^-_{L}\tilde{e}^-_{L} \to e^- e^- jjjj \not\!\!p_T$)~[fb]}}}
\put(83,-3){\makebox(0,0)[br]{{$\sqrt{s}$~[GeV]}}}
\end{picture}
\end{center}
\caption[]{Threshold behaviour of the process in 
  $e^-e^- \to \tilde{e}^-_{L} \tilde{e}^-_{L} \to e^- e^- j j j j p_T
  \hspace{-3.5mm} /$ \, for $P_{e_1}=-0.8$ and $P_{e_2}=-0.8$.  
  ISR corrections and
  beamstrahlung are included. The error bars show the statistical
  error for $\mathcal{L}=1$ fb$^{-1}$. }
\label{the-e-2}
\end{figure}

The process $e^-e^- \to \tilde{e}^-_{L}\tilde{e}^-_{L} \to e^-e^-
j j j j p_T\hspace{-3.5mm} /$ \, leads to  
larger cross sections compared the corresponding process in
$e^-e^+$ annihilation as can be seen in 
\fig{the-e-2}. The cross section at threshold for polarizations
$P_{e_1}=-0.8$ and $P_{e_2}=-0.8$ 
rises much steeper and the mass resolution therefore is much
better than in $e^+e^-$ annihilation. With the effective luminosity 
$\mathcal{L}=1$~fb$^{-1}$ 
the left selectron mass
$m_{\tilde{e}_L}=179.3\pm 0.28$ GeV  can be determined precisely. 
For completeness we note that one obtains in the process
$e^-e^- \to \tilde{e}^{-}_{R}\tilde{e}^{-}_{L} \to  e^-e^-jj\not\!\!p_T$
a mass resolution of $m_{\tilde{e}_L}=179.3\pm 1.1$ GeV. 

\subsection{Comments on scenario II}

The precision for measuring the masses in $e^+e^-$ annihilation as well
as in $e^- e^-$ scattering
depends on the scenario. Since the rate for $e^+e^-\to \tilde{e}_R^+
\tilde{e}_R^-$ is about a factor four 
smaller in scenario (II) the accuracy for $\Delta m_{\tilde{e}_R}$
decreases by about an order of magnitude. In case of
$e^- e^-$ scattering a similar precision is expected for both scenarios. 
 The rates for the mixed
pair $\tilde{e}_R^\pm \tilde{e}_L^-$ are nearly the same as in
scenario (I).  However, the branching ratio for $\tilde{e}_L\to e
\tilde{\chi}^0_2$ is two orders of magnitude smaller, see
Table~\ref{tab_BR}.  Therefore, any conclusions without a detailed
Monte Carlo study, which is beyond the scope of this paper, are very
difficult.

\section{The weak quantum numbers of the selectrons}
\label{sec:4}

Supersymmetry associates to the two chirality states $e^-_L$, $e^-_R$
of the electrons left and right scalar partners $\tilde{e}^-_L$,
$\tilde{e}^-_R$.  In order to test the concept of supersymmetry it is
important to test the weak quantum numbers $R,$ $L$ of the selectrons
produced in $e^+e^-$ annihilation.  For these tests an $e^+e^-$ collider
is better suited than an $e^-e^-$ collider because in the latter one only
the negative charged selectrons can be probed. 
Note, however, that due to CPT the
antiparticles $\tilde{e}^+_L$ and $\tilde{e}^+_R$ are the scalar
partners of $e^+_R$ and $e^+_L$. 

Due to the small electron mass one expects no mixing of the
electroweak eigenstates $\tilde{e}_L$ and $\tilde{e}_R$. The
possibility of a non--vanishing selectron mixing will be discussed in
Sect.~6.  Only at the $e \tilde{e} \tilde{\chi}^0_i$ vertex the chiral
quantum number $L$, $R$ of the electron is uniquely related to its
scalar partner. These vertices appear only in the t--channel but not
in the s--channel. In order to separate these channels the use of both beam
polarizations is absolutely needed. In particular for electrons and
positrons with the same helicity (chirality) only t--channel exchange
of neutralinos contribute so that in the processes $e^+_L e^-_L\to
\tilde{e}^+_R
\tilde{e}^-_L$ and $e^+_R e^-_R \to
\tilde{e}^+_L \tilde{e}^-_R$ they are directly coupled to their scalar
partners.  Therefore, the $R$, $L$ quantum number are correlated to
the charge of the produced selectrons.  In case the electron and positron
have different helicities, $RR$ or $LL$ pairs are produced in the
t--channel, $e^+_L e^-_R\to \tilde{e}^+_R \tilde{e}^-_R$ and $e^+_R
e^-_L\to
\tilde{e}^+_L \tilde{e}^-_L$. However, in this configuration one
obtains also both pairs $RR$ and $LL$ from the s--channel exchange.

\begin{figure}
\setlength{\unitlength}{1mm}
\begin{center}
\begin{picture}(83,70)
\put(-5,-1){\mbox{\epsfig{figure=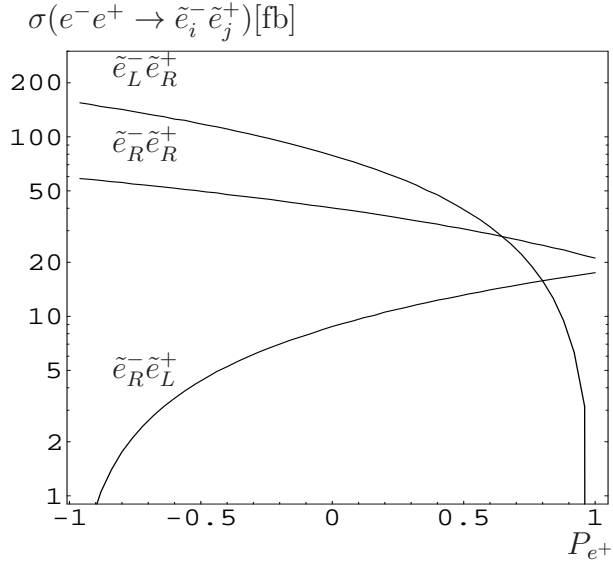,height=6.7cm,width=8.cm}}}
\put(-2,66){\makebox(0,0)[bl]{{$\sigma(e^- e^+\to \selm{i} \selp{j})$[fb]}}}
\put(18,60){\makebox(0,0)[br]{{$\selm{L} \selp{R}$}}}
\put(18,50){\makebox(0,0)[br]{{$\selm{R} \selp{R}$}}}
\put(18,20){\makebox(0,0)[br]{{$\selm{R} \selp{L}$}}}
\put(76,-3){\makebox(0,0)[br]{{$P_{e^+}$}}}
\end{picture}
\end{center}
\caption[]{Production cross sections as a function of $P_{e^+}$ for
    $\sqrt{s} = 350$~GeV, $P_{e^-}=-0.8$, $m_{\tilde{e}_R}=137.7$ GeV,
  $m_{\tilde{e}_L}=179.3$ GeV, $M_2=156$~GeV, $\mu=316$~GeV and
  $\tan\beta=3$. 
  ISR corrections and beamstrahlung are included. 
          }
\label{fig:QN_350}
\end{figure}

For the TESLA design \cite{TDR} a maximal electron (positron)
polarization of $|P_{e^-}|=0.8$ ($|P_{e^+}|=0.6$) is proposed.
Therefore we study the extent to which the test of the $L$, $R$
quantum numbers is possible with partially polarized beams.  In
\fig{fig:QN_350} we show the cross sections for $e^+ e^-\to
\tilde{e}^+_{L,R} \tilde{e}^-_{L,R}$ for $P_{e^-}=-0.8$ as a function
of the positron polarization at $\sqrt{s}=350$~GeV. The SUSY
parameters are specified as in scenario (I).  For positron
polarization $P_{e^+}\lsim 0.5$ the production cross section for
$\tilde{e}^-_L\tilde{e}^+_R$ is the largest one.  We give in
Table~\ref{tab_1} the production cross section for different centre of
mass energies for $P_{e^-}=-0.8$ and $P_{e^+}=-0.6$. The ratio
$r=\sigma(e^+ e^-
\to\tilde{e}^-_L\tilde{e}^+_R) / [\sigma(e^+ e^-
\to\tilde{e}^-_R\tilde{e}^+_R) + \sigma(e^+ e^-
\to\tilde{e}^-_R\tilde{e}^+_L) + \sigma(e^+ e^-
\to\tilde{e}^-_L\tilde{e}^+_L)]$ is larger for $\sqrt{s} = 350$~GeV
($r \simeq 2.4$) than for $\sqrt{s} = 500$~GeV ($r \simeq 1.2$).  The
decrease is due to kinematical effects. This ratio becomes larger with
increasing electron and positron polarization: It would be 7.3 if one
could polarize both beams with 90\%.
We want to stress again that for this investigation polarized positrons
are essential in order to suppress the s--channel
contribution.

\begin{table}
\begin{center}
\begin{tabular}{ccccc}
\hline\noalign{\smallskip}
  &  \multicolumn{4}{c}{$\sqrt{s}$~[GeV]} \\
$\sigma(\tilde{e}^+\tilde{e}^-)$~[fb] &
   350 & 400 & 450 & 500  \\
\noalign{\smallskip}\hline\noalign{\smallskip}
$\tilde{e}^+_L \tilde{e}^-_L$ & 0 & 14 & 43 & 72 \\
$\tilde{e}^+_R \tilde{e}^-_R$ & 52 & 88 & 114 & 130 \\
$\tilde{e}^+_R \tilde{e}^-_L$ & 125 & 205 & 234 & 239\\
$\tilde{e}^+_L \tilde{e}^-_R$ & 3.5 & 5.7 & 6.5 & 6.6 \\ 
\noalign{\smallskip}
\end{tabular}
\end{center}
\caption{Cross sections 
$\sigma(e^+ e^-\to \tilde{e}^+_{L,R}\tilde{e}^-_{L,R})$~[fb] for
$P_{e^-}=-0.8$ and $P_{e^+}=-0.6$ and different $\sqrt{s}$.
ISR and beamstrahlung are included.
The SUSY parameters are chosen as in the reference scenario (I). 
\label{tab_1}}
\end{table}

For $e^+_R e^-_R$ it would be in principle  possible to separate
the pair $\tilde{e}^+_L\tilde{e}^-_R$. 
In case of $P_{e^-}=0.8$, $P_{e^+}=0.6$ the cross section for
$e^+ e^-\to \tilde{e}^-_R \tilde{e}^+_R$ is larger than for 
$P_{e^-}=-0.8$, $P_{e^+}=-0.6$ implying a smaller ratio 
$\sigma(e^+ e^-\to \tilde{e}^-_R \tilde{e}^+_L)/
(\sigma(e^+ e^-\to \tilde{e}^-_R \tilde{e}^+_R)
+\sigma(e^+ e^-\to \tilde{e}^-_L \tilde{e}^+_R)
+\sigma(e^+ e^-\to \tilde{e}^-_L \tilde{e}^+_L)) \simeq 1.13$ for 
$\sqrt{s}=350$~GeV. 
Thus, the precision of the determination of the weak Quantum numbers is in
this case significantly worse compared the case $\tilde{e}^+_R\tilde{e}^-_L$
discussed above.
\begin{figure*}[ht]
\setlength{\unitlength}{1mm}
\begin{center}
\begin{picture}(170,121)
\put(0,-1){\mbox{\epsfig{figure=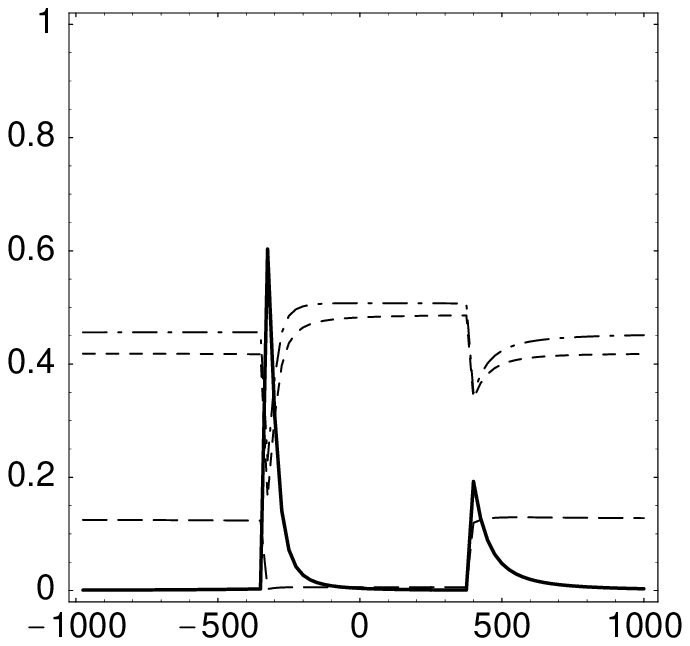,height=5.5cm,width=8.cm}}}
\put(0,61){\mbox{\epsfig{figure=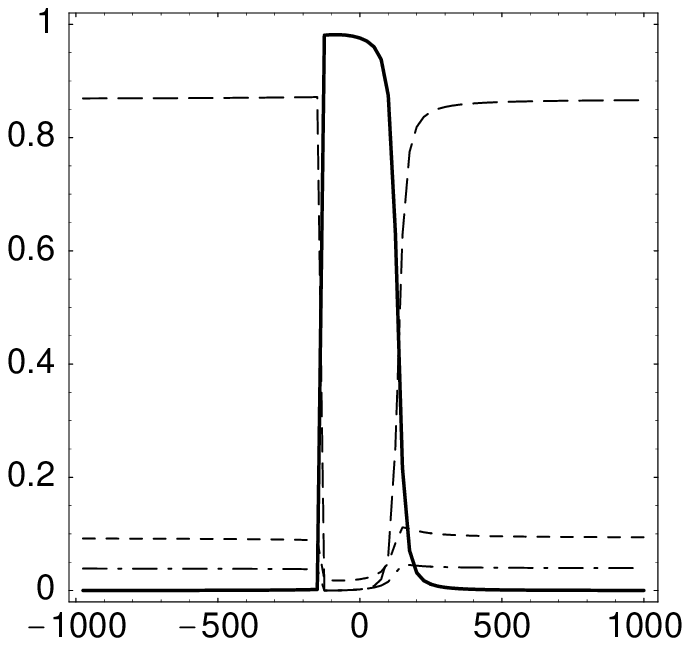,height=5.5cm,width=8.cm}}}
\put(90,-1){\mbox{\epsfig{figure=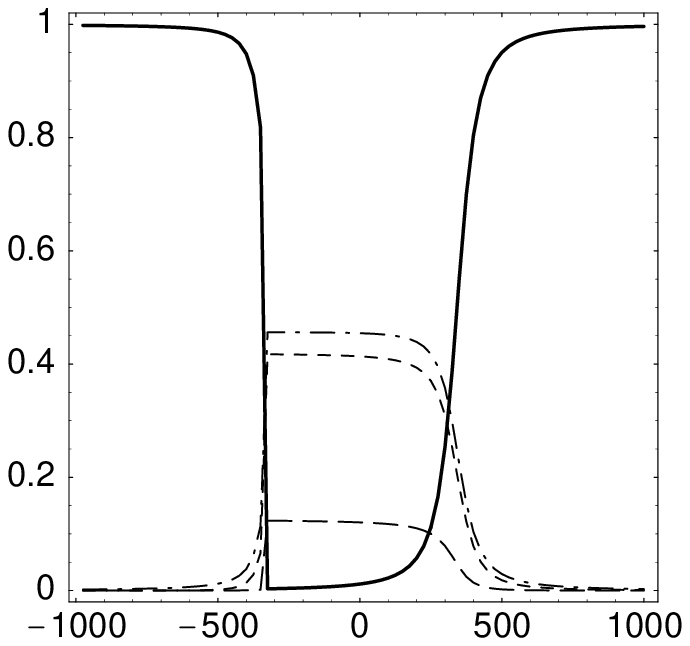,height=5.5cm,width=8.cm}}}
\put(89,61){\mbox{\epsfig{figure=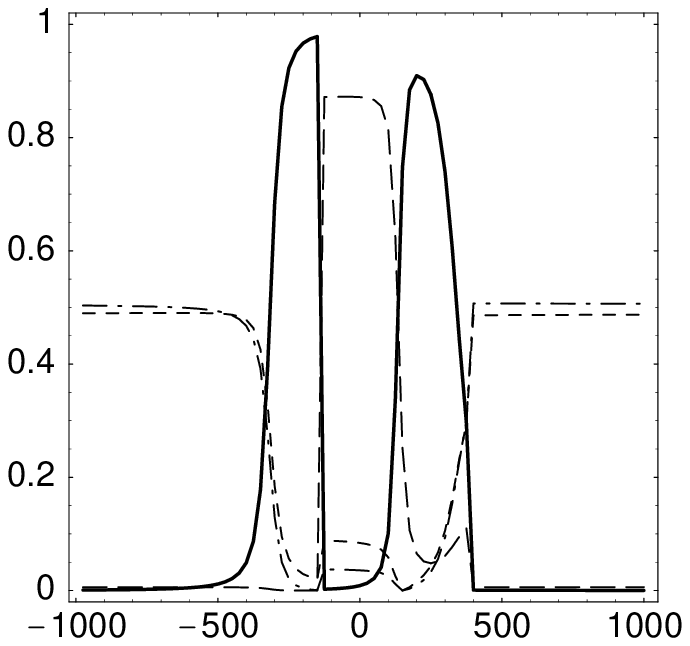,height=5.5cm,width=8.cm}}}
\put(-5,116){\makebox(0,0)[bl]{{{\bf a)}}}}
\put(0,115){\makebox(0,0)[bl]{{$|N_{1j}|^2$}}}
\put(84,59){\makebox(0,0)[br]{{$M_1$~[GeV]}}}
\put(85,116){\makebox(0,0)[bl]{{{\bf b)}}}}
\put(89,115){\makebox(0,0)[bl]{{$|N_{2j}|^2$}}}
\put(172,59){\makebox(0,0)[br]{{$M_1$~[GeV]}}}
\put(-5,54){\makebox(0,0)[bl]{{{\bf c)}}}}
\put(0,53){\makebox(0,0)[bl]{{$|N_{3j}|^2$}}}
\put(84,-3){\makebox(0,0)[br]{{$M_1$~[GeV]}}}
\put(85,54){\makebox(0,0)[bl]{{{\bf d)}}}}
\put(89,53){\makebox(0,0)[bl]{{$|N_{4j}|^2$}}}
\put(172,-3){\makebox(0,0)[br]{{$M_1$~[GeV]}}}
\end{picture}
\end{center}
\caption[]{a) $|N_{1j}|^2$, b) $|N_{2j}|^2$, c) $|N_{3j}|^2$, and
d) $|N_{4j}|^2$ as a function of $M_1$ for $M_2=152$~GeV, $\mu=316$~GeV and 
$\tan\beta=3$. The graphs correspond to the following components:
full line $\tilde{B}$, long dashed line $\tilde{W_3}$, dashed line
$\tilde{H}_d^0$ and long short dashed line $\tilde{H}_u^0$.}
\label{fig:Chi0Nature}
\end{figure*}

\section{$M_1$ Dependence}
\label{sec:5}

The cross sections for production of selectron pairs
in $e^+ e^-$ annihilation and in $e^-e^-$ scattering show a significant 
dependence on the
gaugino mass parameter $M_1$ \cite{WER}. In the following
we show that the measurement of the cross sections for production
and subsequent leptonic decay of the selectrons with polarized beams 
is useful for the determination of $M_1$.

We display therefore in 
\fig{fig:Chi0Nature} the content 
$|N_{ij}|^2$ of the neutralino mass eigenstates $\tilde{\chi}^0_i$ in
the basis $\tilde B$, $\tilde W_3$, $\tilde{H}^0_d $ and
$\tilde{H}^0_u$ \cite{Haber:1985rc} as a function of $M_1$ fixing the
other parameters as in scenario (I).   For
$|M_1| \lsim 150$~GeV the lightest neutralino is bino-like, whereas
for $|M_1| \gsim 150$~GeV it is mainly wino-like. For $|M_1| \lsim
150$~GeV the second lightest neutralino is a wino whereas in the
region 150~GeV $\lsim |M_1|\lsim 400$~GeV it is mainly a bino and for
larger values it is higgsino-like. The neutralino $\tilde{\chi}_3^0$
is always higgsino like and the heaviest is higgsino-like for
$|M_1|\lsim 350$~GeV and bino-like for larger values of $|M_1|$.
For $M_1 \lsim -200$~GeV $\tilde{\chi}^{\pm}_1$ becomes lighter than
$\tilde{\chi}^0_1$ and is thus theoretically excluded.
\begin{figure}[ht]
\setlength{\unitlength}{1mm}
\begin{center}
\begin{picture}(83,57)
\put(0,-15){\includegraphics{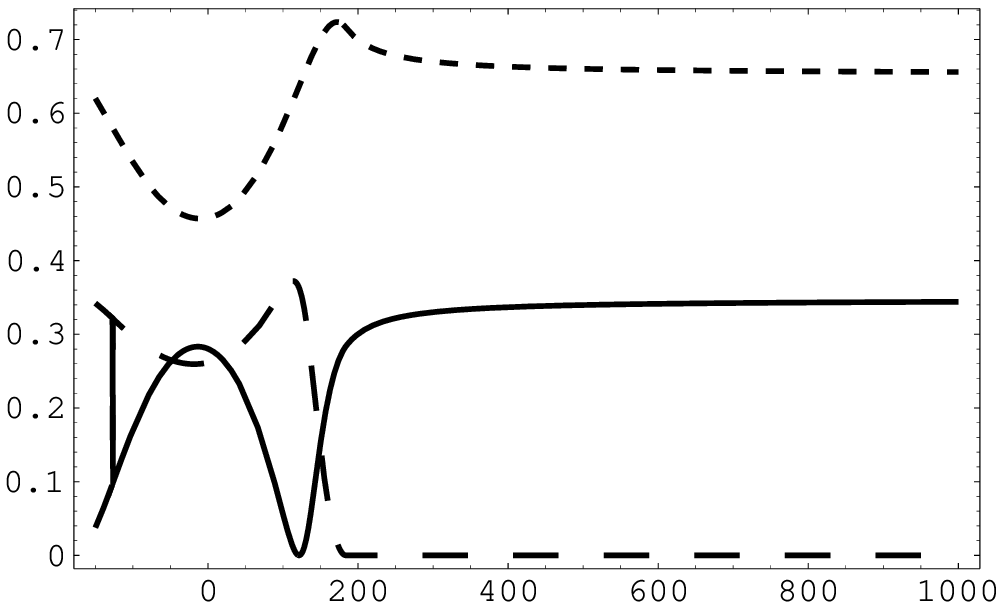}}
\put(0,52){\makebox(0,0)[bl]{{BR$(\tilde e_L)$}}}
\put(83,-2.5){\makebox(0,0)[br]{{$M_1$~[GeV]}}}
\end{picture}
\end{center}
\caption[]{Branching ratios of $\tilde e_L$ as a function of $M_1$ for
   $m_{\tilde{e}_L}=179.3$ GeV, $M_2=152$~GeV
   , $\mu=316$~GeV and $\tan\beta=3$.  The graphs correspond to:
    full line $BR(\tilde{e}_L\rightarrow\tilde{\chi}_1^0e^{\pm})$,
    long dashed line $BR(\tilde{e}_L\rightarrow\tilde{\chi}_2^0e^{\pm})$
    and dashed line $BR(\tilde{e}_L\rightarrow\tilde{\chi}_1^{\pm}\nu_e)$.}
\label{fig:SleptDecay}
\end{figure}

Let us now briefly discuss the $M_1$ dependence of the 
selectron decays. The right selectron decays mainly into the
kinematically accessible neutralino with the largest bino-component.
For the chosen parameters the preferred decay is that
into the lightest neutralino and an electron. For the left
selectron more channels are open with the BR's displayed in
\fig{fig:SleptDecay}. It decays mainly into a chargino and a neutrino.
The second important decay mode is into that kinematically accessible
neutralino with the largest wino-content.

\begin{figure*}
\setlength{\unitlength}{1mm}
\begin{center}
\begin{picture}(160,72)
\put(0,-1){\mbox{\epsfig{figure=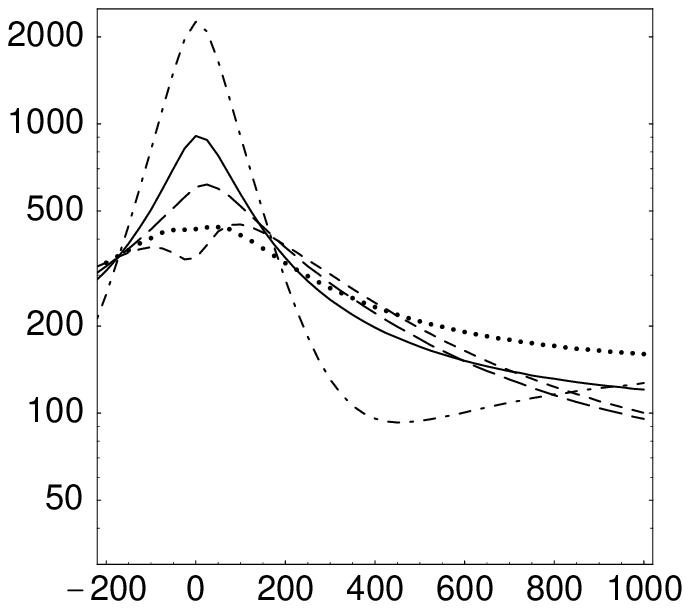,height=7.cm,width=7.cm}}}
\put(-4,67){\makebox(0,0)[bl]{{{\bf a)}}}}
\put(1,66){\makebox(0,0)[bl]{{
 $\sigma$~[fb]}}}
\put(20,66){\makebox(0,0)[bl]{{
 $e^+ e^- \to \sum_{i,j=L,R} \selm{i} \selp{j}$}}}
\put(70,0){\makebox(0,0)[br]{{$M_1$~[GeV]}}}
\put(80,-1){\mbox{\epsfig{figure=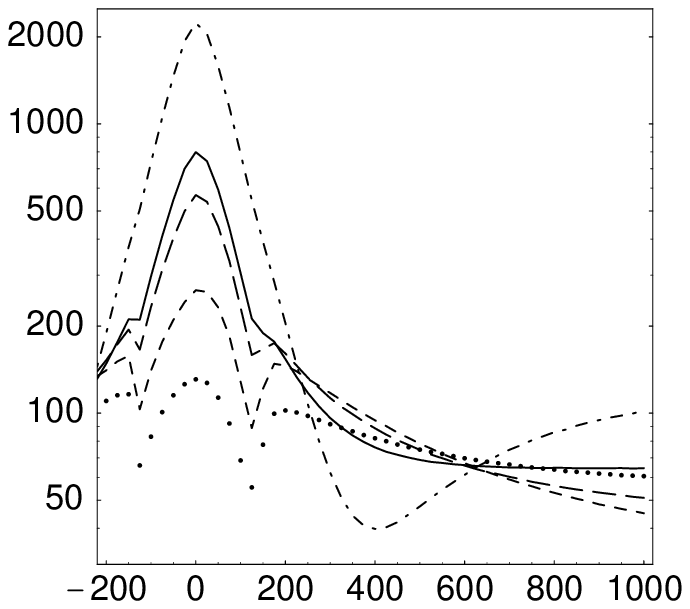,height=7.cm,width=7.cm}}}
\put(77,67){\makebox(0,0)[bl]{{{\bf b)}}}}
\put(81,66){\makebox(0,0)[bl]{{
 $\sigma$~[fb]}}}
\put(93,66){\makebox(0,0)[bl]{{
 $e^+ e^- \to \sum_{i,j=L,R} \selm{i}\selp{j} \to e^+ e^-  
 \slash \hspace{-1.5mm} p_T$}}}
\put(150,0){\makebox(0,0)[br]{{$M_1$~[GeV]}}}
\end{picture}
\end{center}
\caption[]{Cross sections for the processes
  $\sigma(e^+ e^- \to \sum_{i,j=L,R} \selm{i} \selp{j})$ 
  ({\bf a})  and 
  $\sigma(e^+ e^- \to \sum_{i,j=L,R} \selm{i}\selp{j} \to e^+ e^-
    \slash \hspace{-1.5mm} p_T)$
  ({\bf b}) as a function of $M_1$ for
   various polarizations. The effects of ISR- and beamstrahlung corrections are
  included. The graphs
   correspond to the following set of polarizations:
   full line $P_{e^-}=0$, $P_{e^+}=0$,
   dashed line $P_{e^-}=-0.8$, $P_{e^+}=-0.6$,
   dashed-dotted line $P_{e^-}=0.8$, $P_{e^+}=-0.6$,
   dotted line $P_{e^-}=-0.8$, $P_{e^+}=0.6$, and
   long dashed line $P_{e^-}=0.8$, $P_{e^+}=0.6$.
          }
\label{fig:M1dependence2}
\end{figure*}

In the following we study for $\sqrt{s}=500$~GeV
the $M_1$ dependence of the total cross section
$\sigma(e^+ e^- \to \sum_{ij} \tilde{e}_i^+ \tilde{e}^-_j)$
($i,j=R,L$)
and the cross section
$\sigma(e^+ e^- \to \sum_{i,j} \tilde{e}_i^+ \tilde{e}^-_j \to e^+ e^- p_T 
\hspace{-3.5mm} / \, \,)$. 
Since the right selectron decays mainly into the kinematically 
accessible neutralino with the largest bino--component,
the direct decay $\tilde{e}^{\pm}_R\to e^{\pm} \tilde{\chi}^0_1$ 
dominates clearly for the chosen parameters.
 In our scenario the most important decay mode of $\tilde e^\pm_L$
is that via the lighter chargino 
$e^{\pm}_L \to \tilde{\chi}^{\pm}_1\nu_e\to e^{\pm}\tilde{\chi}^0_1 2 \nu_e$.
The second important decay mode is that via the kinematically accessible
neutralino with the largest wino--component. For large $|M_1|$ values
this is the direct decay $\tilde{e}^{\pm}_L \to \tilde{\chi}^0_1 e^{\pm}$
into an electron and the LSP, whereas for smaller values of $M_1$
this decay competes with the cascade decay 
$\tilde{e}^{\pm}\to \tilde{\chi}^0_2 e^{\pm}\to e^{\pm} \tilde{\chi}^0_1\nu
\bar{\nu}$. 
The branching ratios depend also on the squark sector. For a squark
mass of 440~GeV one obtains in our scenario (I) a branching ratio
$BR(\tilde{e}^{\pm}_L \to e^{\pm} \tilde{\chi}^0_1 2 \nu)=0.14$.  We
neglect in the following discussion its $M_1$--dependence because the
variation of the chargino/neutralino branching ratios due to $M_1$ can
be compensated by a change in the squark and/or Higgs sector.
In \fig{fig:M1dependence2} we show the $M_1$ dependence of {\bf a)}
the cross section for selectron production summed over all final
states and {\bf b)} the cross section for the decay leptons.  The
influence of ISR and beamstrahlung is different for the divers
selectron pairs due to different kinematics. In case of $\tilde{e}_L
\tilde{e}_L$ the cross section is decreased compared to the
tree--level value whereas for the other selectron final state it is
increased for $\sqrt{s}=500$~GeV. For larger values of $\sqrt{s}$ also
the $\tilde{e}_L \tilde{e}_L$ cross section increases compared to the
tree-level. The $M_1$ dependence of the cross sections for the different
selectron final states remains, however, unchanged. Since the ratios of 
production cross sections for the different selectron pairs depends
on the beam polarization the influence of ISR and beamstrahlung
modifies this polarization effect.

The $M_1$ dependence of both the cross section for selectron production,
\fig{fig:M1dependence2}a, and for the decay electrons,
\fig{fig:M1dependence2}b, depends on the beam polarization and is
strongest for $P_{e^-}=+0.8$, $P_{e^+}=-0.6$. In this case mainly
$\tilde{e}^+_R \tilde{e}^-_R$ pairs are produced, 
cf.~\fig{fig:M1dependence2}, which couple only to the bino component
of the exchanged neutralinos. Since with increasing $|M_1|$ also the mass
of the bino--like neutralino increases, the cross section decreases
up to $|M_1|\sim 400$ GeV. The reason for the increase of the cross section
for $|M_1|\ge 400$~GeV is the weaker $M_1$ dependence of the contribution 
from t--channel exchange compared to that of the destructive interference 
between s-- and t--channel.
The kink near $M_1=150$~GeV in the cross section for the decay electrons,
\fig{fig:M1dependence2}b, is due to the change of the mixing
character of the neutralinos, cf.~\fig{fig:Chi0Nature}, which 
leads to a strong $M_1$ dependence of the branching ratios shown
in \fig{fig:SleptDecay}.

\begin{figure*}
\setlength{\unitlength}{1mm}
\begin{center}
\begin{picture}(160,70)
\put(0,-1){\mbox{\epsfig{figure=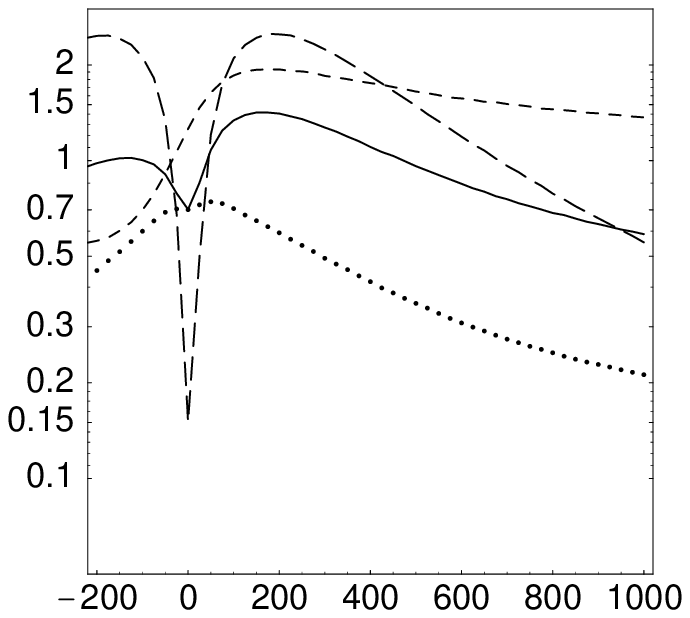,height=6.8cm,width=7.cm}}}
\put(-4,67){\makebox(0,0)[bl]{{{\bf a)}}}}
\put(1,66){\makebox(0,0)[bl]{{
 $\sigma$~[pb]}}}
\put(20,66){\makebox(0,0)[bl]{{
 $e^- e^- \to \sum_{i,j=L,R} \selm{i} \selm{j}$}}}
\put(70,0){\makebox(0,0)[br]{{$M_1$~[GeV]}}}
\put(80,-1){\mbox{\epsfig{
          figure=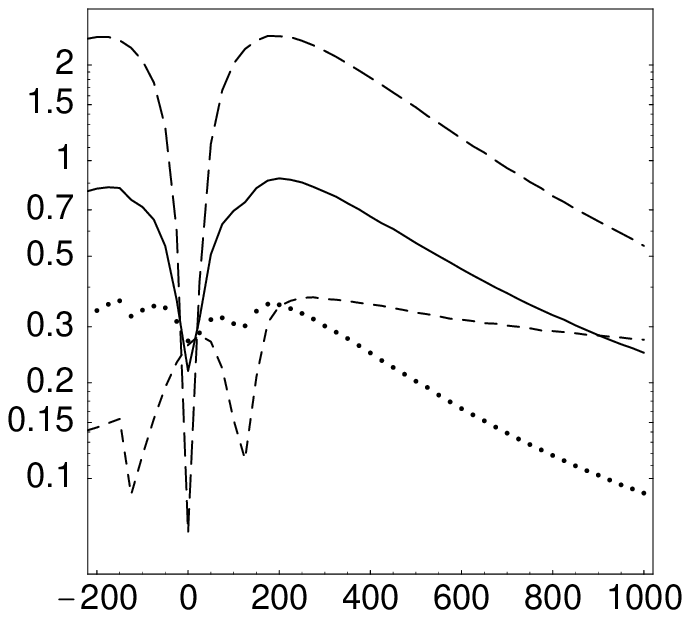,height=6.8cm,width=7.cm}}}
\put(77,67){\makebox(0,0)[bl]{{{\bf b)}}}}
\put(81,66){\makebox(0,0)[bl]{{
 $\sigma$~[pb]}}}
\put(93,66){\makebox(0,0)[bl]{{
 $e^- e^- \to \sum_{i,j=L,R} \selm{i}\selm{j} \to e^- e^-  
 \slash \hspace{-1.5mm} p_T$}}}
\put(150,0){\makebox(0,0)[br]{{$M_1$~[GeV]}}}
\end{picture}
\end{center}
\caption[]{Cross sections for the processes
  $\sigma(e^- e^- \to \sum_{i,j=L,R} \selm{i} \selm{j})$ 
  ({\bf a})  and 
  $\sigma(e^- e^- \to \sum_{i,j=L,R} \selm{i}\selm{j} \to e^- e^-
    \slash \hspace{-1.5mm} p_T)$
  ({\bf b}) as a function of $M_1$ for
   various polarizations. The effects of ISR- and beamstrahlung corrections are
  included. The graphs
   correspond to the following set of polarizations:
   full line $P_{e_1}=0$, $P_{e_2}=0$,
   dashed line $P_{e_1}=-0.8$, $P_{e_2}=-0.8$,
   dotted line $P_{e_1}=-0.8$, $P_{e_2}=0.8$, and
   long dashed line $P_{e_1}=0.8$, $P_{e_2}=0.8$.
          }
\label{fig:M1dependenceEmEm}
\end{figure*}

In case of $e^- e^-$ the polarization dependence of the cross sections
is more pronounced for large values of $|M_1|$ than in case of $e^+e^-$
as can be seen in \fig{fig:M1dependenceEmEm}. This is due to the fact that the
s-channel and therefore the destructive interference is absent.
Also the cross sections are significantly larger for large $|M_1|$  in case of 
of $e^- e^-$ compared to $e^+e^-$ for the same reason.
It is obvious from Figs.~\ref{fig:M1dependence2} and
\ref{fig:M1dependenceEmEm} that measuring the selectron production
cross section using various combinations of electron/positron
polarization is a useful tool to cross check the determination of $M_1$
from other measurements \cite{CKMZ} if not providing the first
measurement by itself. This might be the case if there are additional
neutralinos, as e.g.~in extended supersymmetric models \cite{NMSSM} or if
R-parity is broken spontaneously \cite{Rparity} with a right-handed
scale in the 100~GeV range. In these models are more than four neutralinos and
thus the observation of four
neutralinos does not imply that one has found a bino-like neutralino.
Moreover, different polarization configurations, in particular in case
of $e^- e^-$, are an useful tool
to determine the relative phase between $M_1$ and $M_2$. Most clearly this can
be seen 
in \fig{fig:M1dependenceEmEm}a for the configurations 
$P_{e_1}=-0.8$, $P_{e_2}=-0.8$ (dashed line) and $P_{e_1}=0.8$, $P_{e_2}=0.8$
(long dashed line).

\section{Non-vanishing selectron mixing}
\label{sec:6}

\begin{figure*}
\setlength{\unitlength}{1mm}
\begin{center}
\begin{picture}(150,70)
\put(2,-1){\mbox{\epsfig{
      figure=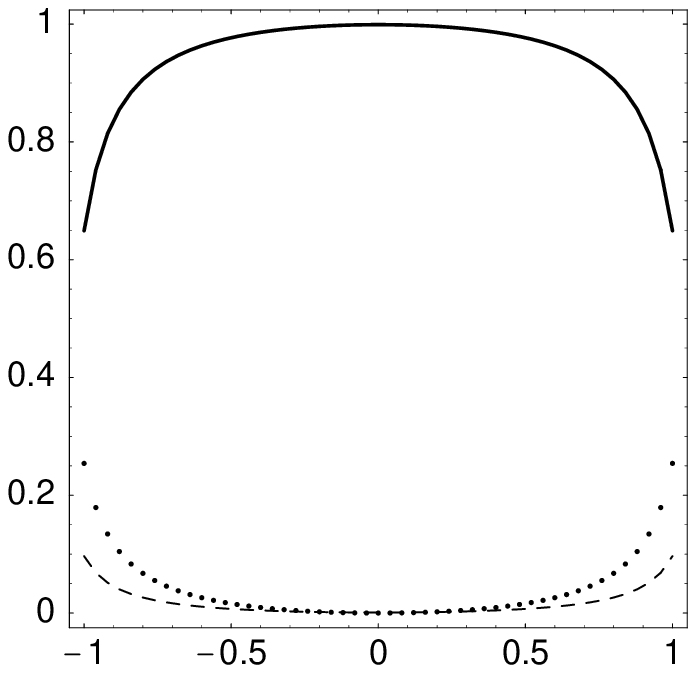,height=6.7cm,width=7.cm}}}
\put(0,67){\makebox(0,0)[bl]{{{\bf a)}}}}
\put(4,66){\makebox(0,0)[bl]{{
 $BR(\selm{1})$}}}
\put(72,-3){\makebox(0,0)[br]{{$\cose$}}}
\put(83,-1){\mbox{\epsfig{
        figure=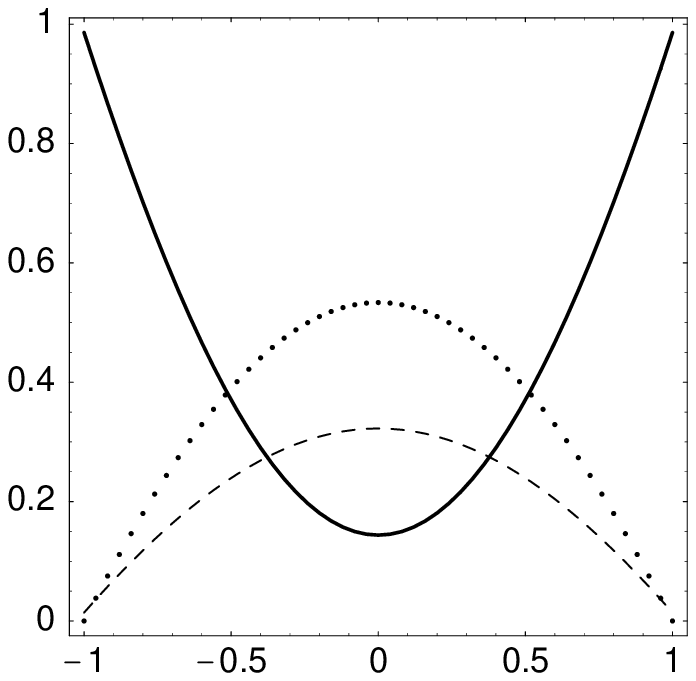,height=6.7cm,width=7.cm}}}
\put(81,67){\makebox(0,0)[bl]{{{\bf b)}}}}
\put(85,66){\makebox(0,0)[bl]{{
 $BR(\selm{2})$}}} 
\put(153,-3){\makebox(0,0)[br]{{$\cose$}}}
\end{picture}
\end{center}
\caption[]{Branching ratios as a function of $\cose$ for
  $\selm{1}$ ({\bf a}) and $\selm{2}$
  ({\bf b}). Masses and gaugino parameters are as in scenario (I).
  The graphs correspond to: $BR(\selm{i} \to e \tilde \chi^0_1)$ full line,
   $BR(\selm{i} \to e \tilde \chi^0_2)$ dashed line, and
   $BR(\selm{i} \to \nu_e \tilde \chi^-_1)$ dotted line.
}
\label{fig:BrSel2}
\end{figure*}

In this section we study  consequences of a non-vanishing selectron mixing,
which although theoretically disfavoured cannot be excluded. This
can be seen by inspecting the mass matrix:
\begin{eqnarray}
{\cal M}^2_{\tilde e} = 
      \left( \begin{array}{cc}
        M^2_L + v^2_1 h^2_e  + D_L & v_1 A_e - \mu h_e v_2  \\
        v_1 A_e - \mu h_e v_2 &  M^2_E + v^2_1 h^2_e  + D_R \end{array} \right)
\end{eqnarray}
with the D-terms
$D_L = (-{\frac{1}{2}} + \sin^2 \theta_W ) \cos(2 \beta)  m^2_Z$ and
$D_R = - \sin^2 \theta_W \cos(2 \beta)  m^2_Z$.
Here $M_L^2$, $M_E^2$, $h_e$, $A_e$, $v_i$ are the left slepton mass parameter,
the right slepton mass parameter, the lepton Yukawa coupling, the trilinear
slepton--Higgs coupling and the vacuum expectation values, respectively.
Motivated by supergravity theories it
is usually assumed that $A_e = h_e \cdot O(100)$~GeV and in this case
the off-diagonal element is negligible. However, this need not to be the
case and it might be that $v_1 A_e$ is large enough to induce a sizable
mixing between left- and right selectrons. 
In the following we use the convention 
$\tilde e_1 = \cose \tilde e_L + \sine \tilde e_R$ and
$\tilde e_2 = -\sine \tilde e_L + \cose \tilde e_R$. 
Let us first
study the dependence of the branching on $\cose$ which
is displayed in \fig{fig:BrSel2}. Here we take 
$m_{\tilde e_1} = 137.7$~GeV, $m_{\tilde e_2} = 179.3$~GeV and the 
gaugino/higgsino parameters as in scenario (I). Note, that for
$\cose=0$ we have exactly the configuration of scenario (I)
whereas for $\cose=\pm 1$ left and right selectrons would exchange
the roles as it might happen for example in theories with extra D-terms
at the unification scale \cite{Kolda:1995iw}.

\begin{figure*}
\setlength{\unitlength}{1mm}
\begin{center}
\begin{picture}(150,70)
\put(2,-1){\mbox{\epsfig{
        figure=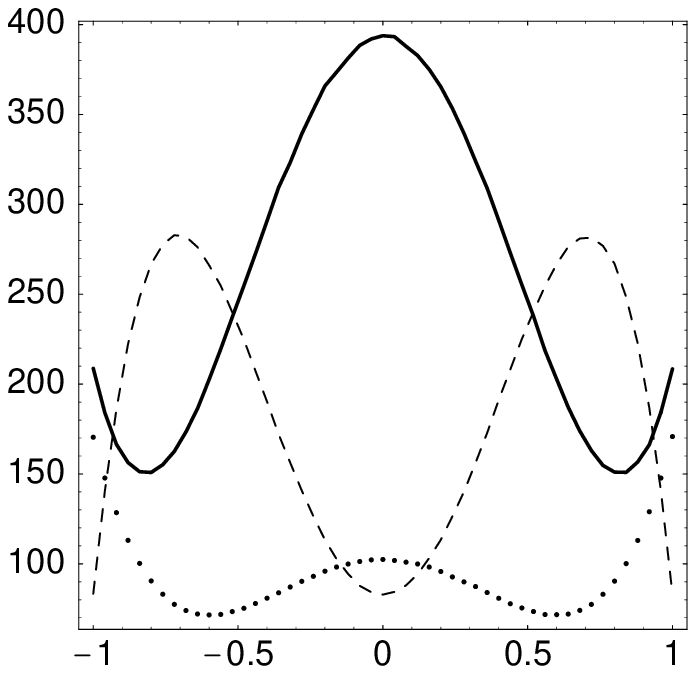,height=6.7cm,width=7.cm}}}
\put(0,67){\makebox(0,0)[bl]{{{\bf a)}}}}
\put(4,66){\makebox(0,0)[bl]{
  {$\sigma(e^+ e^-\to \tilde{e}^-_{i} \tilde{e}^+_{j})$~[fb]}}}
\put(45,55){\makebox(0,0)[br]{{$\tilde e^+_1 \tilde e^-_1$}}}
\put(25,44){\makebox(0,0)[br]{{$\tilde e^+_2 \tilde e^-_1$}}}
\put(25,9){\makebox(0,0)[br]{{$\tilde e^+_2 \tilde e^-_2$}}}
\put(72,-3){\makebox(0,0)[br]{{$\cose$}}}
\put(83,-1){\mbox{\epsfig{
        figure=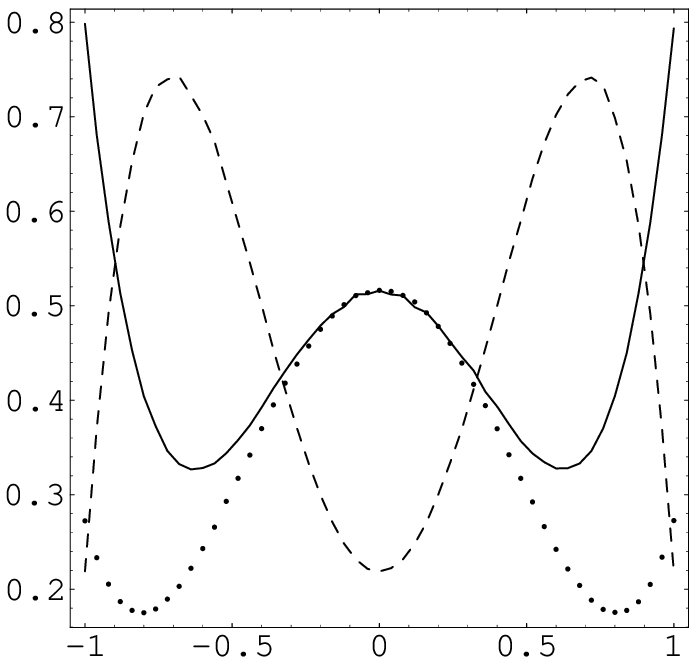,height=6.7cm,width=7.cm}}}
\put(81,67){\makebox(0,0)[bl]{{{\bf b)}}}}
\put(85,66){\makebox(0,0)[bl]{{
 $\sigma(e^- e^-\to \tilde{e}^-_{i} \tilde{e}^-_{j})$~[pb]}}} 
\put(105,31){\makebox(0,0)[br]{{$\tilde e^-_1 \tilde e^-_1$}}}
\put(105,58){\makebox(0,0)[br]{{$\tilde e^-_2 \tilde e^-_1$}}}
\put(102,11){\makebox(0,0)[br]{{$\tilde e^-_2 \tilde e^-_2$}}}
\put(153,-3){\makebox(0,0)[br]{{$\cose$}}}
\end{picture}
\end{center}
\caption[]{Production cross sections for selectrons as a function of 
  $\cose$ at ({\bf a}) an $e^+ e^-$ collider and
  ({\bf b}) an $e^- e^-$ collider for $\sqrt{s} = 500$~GeV and unpolarized 
  beams. ISR and beamstrahlung are included. We take the selectron masses
  and the gaugino/higgsino parameters as in scenario (I). The full line
  shows the cross section for pair production of the lighter selectrons,
  the dashed line a lighter and a heavier selectron, and the dotted line
  a pair of heavier selectrons.
}
\label{fig:Prod500}
\end{figure*}

In \fig{fig:BrSel2}a) we display the branching ratios for the lighter
selectron $\selm{1}$ as a function of $\cose$.  It decays
mainly into the lightest neutralino even for $|\cose|=1$ where
it is a pure left-selectron. This is simply due to
kinematics. Therefore, for $|\cose|\lsim 0.4$ it might be
difficult to decide whether the selectrons mix or not if one looks
only at the branching ratios of the lighter selectron. In case of the
heavier selectron $\selm{2}$ the situation is in general much clearer
as can be seen in \fig{fig:BrSel2}b) because the various decay
channels have sufficient phase space. For $|\cose|\lsim 0.4 -
0.5$ the decay into the lighter chargino (dotted line) dominates
followed by the decay into the wino-like neutralino $\tilde{\chi}^0_2$
(dashed line). For
$|\cose|\gsim 0.5$ the decay into the lighter neutralino 
$\tilde{\chi}^0_1$
dominates which is due to kinematics and due to the fact that the
right-component of the selectron grows.  Note that these features are
quite insensitive to $\tan \beta$ because even for $\tan \beta =50$
the electron Yukawa coupling is still negligible (contrary to the case
of staus as discussed e.g.~in \cite{Bartl}).

In \fig{fig:Prod500} we show the dependence of the cross sections
on $\cose$ for ({\bf a}) an $e^+ e^-$ collider and ({\bf b}) an 
$e^- e^-$ collider at $\sqrt{s} = 500$~GeV with unpolarized beams.  As can
be seen there is a pronounced dependence on $\cose$ at both collider
types. Note that the production cross section for $\selm{1} \selp{2}$
is in general larger than that for the pair production in case of
large mixing, $|\cose| \simeq 1/\sqrt{2}$. Clearly the ratios of the
various production channels are changed with varying beam polarization
similarly as discussed in Sect.~\ref{sec:2}. As in the case of the decays the
results for selectron production are quite insensitive to $\tan\beta$.

\begin{figure}
\setlength{\unitlength}{1mm}
\begin{center}
\begin{picture}(83,70)
\put(0,-2){\mbox{\epsfig{figure=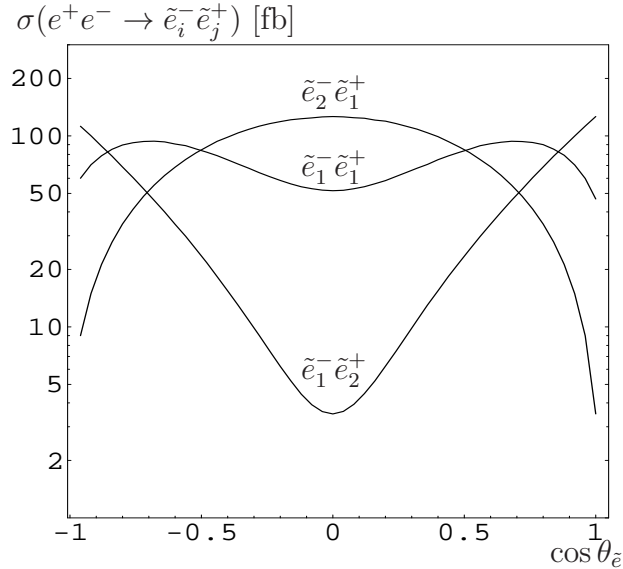,height=7.cm,width=8.cm}}}
\put(0,67){\makebox(0,0)[bl]{{
   $\sigma(e^+ e^-\to \tilde{e}^-_{i} \tilde{e}^+_{j})$~[fb]}}}
\put(48,58){\makebox(0,0)[br]{{$\selm{2} \selp{1}$}}}
\put(48,48){\makebox(0,0)[br]{{$\selm{1} \selp{1}$}}}
\put(48,21){\makebox(0,0)[br]{{$\selm{1} \selp{2}$}}}
\put(82,-3){\makebox(0,0)[br]{{$\cose$}}}
\end{picture}
\end{center}
\caption[]{Production cross sections as a function of $\cose$ for
    $\sqrt{s} = 350$~GeV, $P_{e^-}=-0.8$,  $P_{e^+}=-0.6$, 
    $m_{\tilde{e}_1}=137.7$ GeV,
  $m_{\tilde{e}_2}=179.3$ GeV and the other SUSY parameters as in scenario (I).
  ISR corrections and beamstrahlung are included. 
          }
\label{fig:QN350}
\end{figure}
We have studied in \sect{sec:4} the possibility to test the weak
quantum numbers $L$, $R$ of the selectrons at an $e^+ e^-$
collider. We have shown that one can single out the $\selm{R}
\selp{L}$ state depending on the polarization of the beams and one
predicts a certain ratio of the various cross sections. In
\fig{fig:Prod500} one sees that there are several values of $\cose$
with identical ratios of cross sections. Therefore
the question arises whether it is possible to
obtain the same polarization dependence of all cross sections 
for mixed selectrons as shown in \fig{fig:QN_350} for the unmixed case.
 As can be seen in \fig{fig:QN350} the case
$|\cose| \lsim 0.1$ might be difficult to distinguish
from the unmixed case. However, for $0.2 \lsim |\cose| \lsim0.9$ 
it should be clear whether selectron mixing exists or not
provided the neutralino sector is known so that one can calculate the
cross sections.

\section{Conclusion}
\label{sec:con}

In this paper we have studied selectron pair production in $e^+ e^-$
annihilation and in $e^- e^-$ scattering with polarized electron and
positron beams including ISR and beamstrahlung.  We have shown that at
both collider modes all cross sections have a pronounced dependence on
beam polarization.  We have compared the possibility to measure the
selectron masses using threshold scans at both colliders. Our results
indicate that in the $e^-e^-$ mode one tenth of the luminosity of the
$e^+e^-$ mode should be sufficient to obtain similar precision in the
mass determination.  Beam polarization is a useful tool to enhance the
precision.  Furthermore one can test the association between chiral
electrons and their scalar superpartners at an $e^+ e^-$ collider if
both beams are simultaneously polarized. We want to stress that in
this case the polarization of the positron beam is indispensable.  We
also have studied the dependence of cross sections and branching
ratios on the gaugino mass parameter $M_1$ which is pronounced at both
colliders. In addition we have shown that the phenomenology of
selectrons change significantly in case a non-vanishing mixing between
left-- and right--selectrons is realized in nature.

\section{Acknowledgements}

C.B.~and W.P.~thank M.~Peskin for discussions on beamstrahlung.
This work has been partly supported by the EU TMR Network
Contract No.~HPRN-CT-2000-00149.
C.B.~and H.F.~are supported by the 'Deutsche
Forschungsgemeinschaft' under contract number Fr 1064/5-1,
W.P.~is supported by `Fonds zur F\"orderung der wissenschaftlichen
Forschung' of Austria, Erwin Schr\"odinger fellowship Nr. J2095, and
partly by the `Schweizer Nationalfonds'.

\begin{appendix}

\section{Lagrangian and couplings}

In this section we list the lagrangian and the couplings for the
calculation of selectron pair poduction. The selectron mass matrix in the
most general form within the MSSM is given by:
\begin{eqnarray}
{\cal M}^2_{\tilde e} = 
      \left( \begin{array}{cc}
        M^2_L + v^2_1 h^2_e  + D_L & v_1 A_e - \mu h_e v_2  \\
        v_1 A_e - \mu h_e v_2 &  M^2_E + v^2_1 h^2_e  + D_R \end{array} \right)
\end{eqnarray}
with the D-terms 
$D_L = (-{\frac{1}{2}} + \sin^2 \theta_W ) \cos(2 \beta)  m^2_Z$ and
$D_R = - \sin^2 \theta_W \cos(2 \beta)  m^2_Z$. The mass eigenstates 
are connected via
\begin{eqnarray}
 \left( \begin{array}{c} \sel{}{1} \\ \sel{}{2} \end{array} \right) &=&
 \left( \begin{array}{cc} \cose & \sine \\ -\sine & \cose \end{array}
 \right)
 \left( \begin{array}{c} \sel{}{L} \\ \sel{}{R} \end{array} \right) 
\end{eqnarray}
with the electroweak eigenstates and we take 
$m_{\tilde{e}_1} < m_{\tilde{e}_2}$.

The relevant parts of the interaction Lagrangian are given by:
\begin{eqnarray}\label{L1}
\mathcal{L}_{e\tilde{e}_k\tilde{\chi}^0_j} & = & 
g f_j^L \bar{e} \left( a_{jk} P_R +  b_{jk} P_L \right)\tilde{\chi}^0_j
   \tilde{e}_{k} + h.c.
\\
\mathcal{L}_{\gamma\tilde{e}\tilde{e}} & = & -i e_e A_\mu \left(
      \tilde{e}^{*}_1 \stackrel{\leftrightarrow}{\partial_\mu} \tilde{e}_1
    +  \tilde{e}^{*}_2 \stackrel{\leftrightarrow}{\partial_\mu} \tilde{e}_2
             \right)
 \label{L3}\\
\mathcal{L}_{Z\tilde{e}\tilde{e}} & = & \label{L2}
   - i g Z_\mu c_{ij}  \tilde e^{*}_i 
             \stackrel{\leftrightarrow}{\partial_\mu} \tilde e_j \\
\mathcal{L}_{\gamma ee} & = & -e_e A_\mu \bar{e} \gamma^\mu e \label{L4}\\
\mathcal{L}_{Zee} & = & -g Z_\mu \bar{e} \gamma^\mu (L_e P_L + R_e P_R) e
    \label{L5}
\end{eqnarray}
In the basis $\tilde{B}$, $\tilde{W}^3$, $\tilde{H}^0_d$,
$\tilde{H}^0_u$,
the couplings are given by
\begin{eqnarray}\label{fkop1}
a_{j1} &=& \cose f_j^L \, , \, a_{j2} = - \sine f_j^L \\
b_{j1} &=& \sine f_j^R \, , \, b_{j2} = \cose f_j^R \\
f_j^L & = & \frac{1}{\sqrt{2}} \left( N_{j2} + \tan \theta_W N_{j1} \right) \\
f_j^R & = & - \sqrt{2}\tan \theta_W N^*_{j1} \\
c_{11} &=& R_e \sinesq + L_e \cosesq \\
c_{12} &=& c_{21} = (L_e - R_e) \sine \cose \\
c_{22} &=& L_e \sinesq + R_e \cosesq \\
L_e &=& - \frac{1}{\cos\theta_W}\left(\frac{1}{2} - \sin^2 \theta_W \right) \\
R_e &=& \sin \theta_W \tan\theta_W
\end{eqnarray}
$e_e$ is the electromagnetic coupling of the electron, respectively.
We have neglected the electron Yukawa coupling. $N_{ij}$ are the
components of the neutralino mixing matrix.

\section{Amplitudes and cross section for selectron production in
$e^+ e^-$ annihilation}

We give the helicity amplitudes and cross section for the process
$e^-(p_1,\lambda_m) e^+(p_2,\lambda_n) \to
  \tilde{e}^-_{i}(p_3)\tilde{e}^+_{j}(p_4)$, where
$\lambda_m$ ($\lambda_n$) denotes the helicity of $e^-$ ($e^+$) and $i,j=1,2$.

The helicity amplitudes $T^{\lambda_m \lambda_n}_{ij}$ are:
\begin{eqnarray}
T_{ij}^{\lambda_m \lambda_n}(\gamma)&=& e^2 \Delta(\gamma)
\delta_{ij} 
\bar{v}(p_2,\lambda_n) (\mbox{$\not\!p$}_4-\mbox{$\not\!p$}_3) u(p_1,\lambda_m)
\nonumber\\
&& \label{app_eq4a}\\
T_{ij}^{\lambda_m \lambda_n}(Z)&=& g^2 c_{ij} \Delta(Z) \nonumber \\
&& \hspace{-5mm}
\times \bar{v}(p_2,\lambda_n) (\mbox{$\not\!p$}_4-\mbox{$\not\!p$}_3) 
(L_{\ell} P_L+R_{\ell} P_R) u(p_1,\lambda_m)  \nonumber\\
&&\label{app_eq4b}\\
T_{ij}^{\lambda_m \lambda_n}(\tilde{\chi}^0_k) &=&
 g^2  \, \Delta(\tilde{\chi}^0_k) \,
\bar{v}(p_2,\lambda_n) (a_{ki} P_R + b_{ki} P_L ) \nonumber \\
&& \hspace{-5mm}
   \times (\mbox{$\not\!p$}_1-\mbox{$\not\!p$}_3 + m_{\tilde{\chi}^0_k}) 
   (a^*_{kj} P_L + b^*_{kj} P_R ) u(p_1,\lambda_m) \nonumber \\
\label{app_eq4c}
\end{eqnarray}
The propagators are: 
\begin{eqnarray}
\Delta(\gamma)&=&\frac{i}{(p_1 + p_2)^2}\\
\Delta(Z)&=&\frac{i}{(p_1 + p_2)^2-m^2_Z+im_Z\Gamma_Z}\\
\Delta(\tilde{\chi}^0_k)&=&
\frac{i}{(p_1 - p_3)^2-m^2_{\tilde{\chi}^0_k}}\,\label{app_eq5a}
\end{eqnarray}

The amplitudes squared are most easily expressed in terms of 
Mandelstam variables: $s=(p_1 + p_2)^2=(p_3 + p_4)^2$,
$t=(p_1 - p_3)^2=(p_2 - p_4)^2$ and $u=(p_1 - p_4)^2=(p_2 - p_3)^2$.
We further introduce four polarization factors:
\begin{eqnarray}
c_{-+} & = & (1-P_{e^-})(1+P_{e^+})\label{app_eq7a}\\
c_{+-} & = & (1+P_{e^-})(1-P_{e^+})\label{app_eq7b}\\
c_{--} & = & (1-P_{e^-})(1-P_{e^+})\label{app_eq7c}\\
c_{++} & = & (1+P_{e^-})(1+P_{e^+})\label{app_eq7d}
\end{eqnarray}
where the first (second) index denotes the sign of the
favoured helicity of $e^-$ 
($e^+$) and $P_{e^{\pm}}$ gives the corresponding
polarization of $e^{\pm}$.
\begin{eqnarray}
|T_{ij}(\gamma)|^2 &=& 4 e^4_e \delta_{ij}  |\Delta(\gamma)|^2
             (c_{-+} + c_{+-}) (u \, t - m^2_i m^2_j)  \\
T_{ij}(\gamma) T_{ij}^*(Z)  &=& 4 e^2_e g^2 \delta_{ij} c_{ij} \Delta(\gamma) \Delta^*(Z)
           (c_{-+} L_e + c_{+-} R_e )  (u \, t - m^2_i m^2_j) \\
T_{ij}(\gamma) T_{ij}^*(\tilde{\chi}^0_k)  &=&
 2 e^2_e g^2 \delta_{ij} \Delta(\gamma)
   \Delta^*(\tilde{\chi}^0_k) 
 \nonumber \\
    && \hspace{-7mm} \times
     (c_{-+} |a_{ki}|^2 + c_{+-} |b_{ki}|^2 )(u \, t - m^2_i m^2_j) \\
|T_{ij}(Z)|^2  &=& 4 g^4 c^2_{ij} |\Delta(Z)|^2
           (c_{-+} L^2_e + c_{+-} R^2_e )    (u \, t - m^2_i m^2_j)\\
T_{ij}(Z) T_{ij}^*(\tilde \chi^0_k)  &=& 2 g^4 c_{ij} \Delta(Z)
   \Delta^*(\tilde \chi^0_k) 
 \nonumber \\
    &&  \hspace{-7mm} \times  
    (c_{-+} a_{ki} a^*_{kj} L_e + c_{+-} b_{ki} b^*_{kj} R_e  )
    (u \, t - m^2_i m^2_j) \nonumber \\ && \\
T_{ij}(\tilde \chi^0_k) T_{ij}^*(\tilde \chi^0_l)  &=& 2 g^4 \Delta(\tilde \chi^0_k)
   \Delta^*(\tilde \chi^0_l)  \nonumber \\
  && \hspace{-9mm} \times 
    \bigg[ \left( c_{++} a_{li} a^*_{ki} b_{lj} b^*_{kj}
                 + c_{--} a_{lj} a^*_{kj} b_{li} b^*_{ki} \right) m_k m_l s
 \nonumber \\
 &&  \hspace{-6mm} + \left( c_{-+} a_{li} a^*_{ki} a_{lj} a^*_{kj}
                 + c_{+-} b_{lj} b^*_{kj} b_{li} b^*_{ki} \right)
 \nonumber \\
 && \times         (u \, t - m^2_i m^2_j)  \bigg]
\end{eqnarray}
Note that the terms proportional to $s$ gives rise to an $\beta$
dependence of the cross section near the thresold whereas terms proportional
to $u \, t - m^2_i m^2_j$ gives rise to the $\beta^3$ dependence of the cross
section near the thresold.

The total cross section is given by:
\begin{eqnarray}
\sigma(e^+ e^- \to \selm{i} \selp{j}) &=&
 \frac{1}{64 \pi\, s }\int d t  \bigg|
\sum_{x=\gamma,Z,\tilde \chi^0_k} T_{ij}(x) \bigg|^2 \, .\label{app_eq10a}
\end{eqnarray}
In the center of mass system one can express the Madelstam variables
in terms of the beam energy $E$ and the angle $\theta$ between the electron and
the selectron $\selm{i}$: $s = 4 E^2$,
$t = (m^2_i + m^2_j - s + \cos \theta \kappa(s,m^2_i,m^2_j))/2$ and
$u = (m^2_i + m^2_j - s - \cos \theta \kappa(s,m^2_i,m^2_j))/2$ with
$\kappa(x,y,z) = \sqrt{(x-y-z)^2 -4 y z}$. From this follows
$u t - m^2_i m^2_j= \sin^2 \theta $  $\times \kappa^2(s,m^2_i,m^2_j)$ 
indicating
the $P$-wave nature of the respective terms. 

\section{Amplitudes and cross section for selectron production in
$e^- e^-$ scattering}

Here we give the helicity amplitudes and cross section for the process
$e^-(p_1,\lambda_m) e^-(p_2,\lambda_n) \to
  \tilde{e}^-_{i}(p_3)\tilde{e}^-_{j}(p_4)$, where
$\lambda_{m,n}$ denotes the helicity of the electrons and $i,j=1,2$.

The helicity amplitudes $T^{\lambda_m \lambda_n}_{ij}$ are given by:
\begin{eqnarray}
T_{ij,t}^{\lambda_m \lambda_n}(\tilde{\chi}^0_k) &=&
 g^2  \, \Delta_t(\tilde{\chi}^0_k) \,
\bar{v}(p_2,\lambda_n) (a^*_{ki} P_L + b^*_{ki} P_R )  \nonumber \\
&& \hspace{-7mm} \times
    (\mbox{$\not\!p$}_1-\mbox{$\not\!p$}_3 + m_{\tilde{\chi}^0_k}) 
   (a^*_{kj} P_L + b^*_{kj} P_R ) u(p_1,\lambda_m) \nonumber \\
&& \\
T_{ij,u}^{\lambda_m \lambda_n}(\tilde{\chi}^0_k) &=&
 g^2  \, \Delta_u(\tilde{\chi}^0_k) \,
\bar{v}(p_2,\lambda_n) (a^*_{kj} P_L + b^*_{kj} P_R ) \nonumber \\
&& \hspace{-7mm} \times
    (\mbox{$\not\!p$}_1-\mbox{$\not\!p$}_4 + m_{\tilde{\chi}^0_k}) 
   (a^*_{ki} P_L + b^*_{ki} P_R ) u(p_1,\lambda_m). \nonumber \\
\end{eqnarray}
The subscripts $t$ and $u$ indicate the $t$-channel and $u$-channel,
respectively. The propagators are given by:
\begin{eqnarray}
\Delta_t(\tilde{\chi}^0_k)&=&
\frac{i}{(p_1 - p_3)^2-m^2_{\tilde{\chi}^0_k}}\,, \\
\Delta_u(\tilde{\chi}^0_k)&=&
\frac{i}{(p_1 - p_4)^2-m^2_{\tilde{\chi}^0_k}}\,.
\end{eqnarray}

For the calculation of the amplitude squared we define the Mandelstam
variables in the same terms of momenta as above. Similarly as above we define
 four polarization factors:
\begin{eqnarray}
c_{-+} & = & (1-P_{e_1})(1+P_{e_2})\\
c_{+-} & = & (1+P_{e_1})(1-P_{e_2})\\
c_{--} & = & (1-P_{e_1})(1-P_{e_2})\\
c_{++} & = & (1+P_{e_1})(1+P_{e_2}),
\end{eqnarray}
where the first (second) index denotes the sign of the
favoured helicity of $e^-(p_1)$ 
($e^-(p_2)$) and $P_{e_{1,2}}$ gives the corresponding
polarization of $e^-(p_{1,2})$.
\begin{eqnarray}
  \label{eq:Mtsq}
T_{ij,t}(\tilde \chi^0_k) T^*_{ij,t}(\tilde \chi^0_l) &=& 
 g^4 \Delta_t(\tilde \chi^0_k)
   \Delta^*_t(\tilde \chi^0_l)  \nonumber \\
 && \hspace{-25mm} \times \bigg[ 
  \left( c_{+-} a_{li} a^*_{ki} b_{lj} b^*_{kj}
      +  c_{-+} a_{lj} a^*_{kj} b_{li} b^*_{ki} \right)
      \left( t \, u - m^2_i m^2_j \right)
  \nonumber \\ 
 && \hspace{-22mm} +  \left( c_{++}  b_{li} b^*_{ki} b_{lj} b^*_{kj}
      + c_{--} a_{li} a^*_{ki} a_{lj} a^*_{kj} \right) m_k m_l s \bigg]
\\
T_{ij,u}(\tilde \chi^0_k) T^*_{ij,u}(\tilde \chi^0_l)  &=& 
  T_{ij,t}(\tilde \chi^0_k) T^*_{ij,t}(\tilde \chi^0_l)
   ( i \leftrightarrow j, t \leftrightarrow u)
\\
T_{ij,t}(\tilde \chi^0_k) T^*_{ij,u}(\tilde \chi^0_l) &=&  g^4 
\Delta_t(\tilde \chi^0_k) \Delta^*_u(\tilde \chi^0_l) 
  \nonumber \\
 && \hspace{-27mm} \times \bigg[-
  \left( c_{-+} a_{lj} a^*_{kj} b_{li} b^*_{ki}
    + c_{+-} a_{li} a^*_{ki} b_{lj} b^*_{kj} \right) (t \, u - m^2_i m^2_j)
  \nonumber \\
 && \hspace{-24mm} +
 \left( c_{--} a_{li} a^*_{ki} a_{lj} a^*_{kj}
      +  c_{++} b_{li} b^*_{ki} b_{lj} b^*_{kj} \right)  m_k m_l \, s \bigg]
\end{eqnarray}
The total cross section is given by:
\begin{eqnarray}
\sigma(e^- e^- \to \selm{i} \selm{j}) &=&  \nonumber \\ 
 && \hspace{-28mm}
 \frac{1}{64 \pi n! \, s }
  \int \hspace{-1mm} d t
   \bigg| \sum_{k=1}^4 T_{ij,t}(\tilde \chi^0_k)+T_{ij,u}(\tilde \chi^0_k)
   \bigg|^2 \, ,
\end{eqnarray}
where $n$ is the number of identical particles in the final state.
The same consideration concering the center of mass system holds as in the
case of $e^+ e^-$.

\end{appendix}

\end{document}